\documentclass[10pt]{iopart}
\usepackage{iopams}
\usepackage{amssymb,bm,latexsym,acronym,float,soul,nicefrac,verbatim,cancel}
\usepackage{multirow,dcolumn,longtable,footnote,appendix,graphicx,color,subfigure}
\usepackage[colorlinks,linkcolor=blue,citecolor=blue,urlcolor=blue]{hyperref}
\usepackage[normalem]{ulem}

\newcommand{\be}{\begin{equation}}
\newcommand{\ee}{\end{equation}}
\newcommand{\bea}{\begin{eqnarray}}
\newcommand{\eea}{\end{eqnarray}}
\newcommand{\cO}{{\cal O}}

\newcommand{\msun}{{\rm~M}_\odot}

\begin{document}

\title{Progress of the TianQin project}

%
%
%
%

\newcommand{\TRC}{MOE Key Laboratory of TianQin Mission, TianQin Research Center for Gravitational Physics $\&$ School of Physics and Astronomy, Frontiers Science Center for TianQin, Gravitational Wave Research Center of CNSA, Sun Yat-sen University (Zhuhai Campus), Zhuhai 519082, China}

\newcommand{\SYSUSPA}{School of Physics and Astronomy, Sun Yat-sen University (Zhuhai Campus), Zhuhai 519082, China}

\newcommand{\SYSUSAI}{School of Artificial Intelligence, Sun Yat-sen University (Zhuhai Campus), Zhuhai 519082, People’s Republic of China}

\newcommand{\HUST}{National Gravitation Laboratory, MOE Key Laboratory of Fundamental Physical Quantities Measurement, and School of Physics, Huazhong University of Science and Technology, Wuhan 430074, China}

\newcommand{\DFH}{DFH Satellite Co., Ltd., Beijing 100094, China}

\newcommand{\KIAA}{Kavli Institute for Astronomy and Astrophysics, Peking University, Beijing 100871, China}

\newcommand{\SAI}{Sternberg Astronomical Institute, M.V. Lomonosov Moscow State University, Moscow 119234, Russia}

\newcommand{\FOA}{National Key Laboratory of Spacecraft Thermal Control of Institute of Spacecraft System Engineering, Beijing 100094, China}
\newcommand{\FOB}{Beijing Institute of Control Engineering, Beijing 100094, China}
\newcommand{\FOC}{Space Star Technology CO., LTD., Beijing 100095, China}
\newcommand{\FOD}{China Academy of Space Technology (Xi' an), Xi' an 710199, Shanxi, China}
\newcommand{\FOE}{Beijing Institute of Space Mechanics \& Electricity, Beijing 100076, China}

%
%
\author{Jun Luo$^1$,
Shaojun Bai$^4$,
Yan-Zheng Bai$^2$,
Lin Cai$^2$,
Hao Dang$^5$,
Qijia Dong$^6$,
Hui-Zong Duan$^1$,
Yuanbo Du$^1$,
Lei Fan$^1$,
Xinju Fu$^7$
% ...
}

\address{$^1$\TRC}
\address{$^2$\HUST}
\address{$^3$\DFH}
\address{$^4$\FOE}
\address{$^5$\FOA}
\address{$^6$\FOC}
\address{$^7$\FOB}
\address{$^8$\SYSUSAI}
\address{$^9$\FOD}
\address{$^{10}$\SAI}
\address{$^{11}$\KIAA}

%
%
\ead{\\
$^\star$meijw@sysu.edu.cn\\
$^\sharp$yexianji@mail.sysu.edu.cn\\
$^\ddagger$zhangxf38@mail.sysu.edu.cn\\
$^\natural$zhouzb@mail.hust.edu.cn}

\vspace{10pt}
\begin{indented}
\item[] \today
\end{indented}

\begin{abstract}
TianQin is a future space-based gravitational wave observatory targeting the frequency window of $10^{-4}$ Hz $\sim 1$ Hz.
A large variety of gravitational wave sources are expected in this frequency band, including the merger of massive black hole binaries, the inspiral of extreme/intermediate mass ratio systems, stellar-mass black hole binaries, Galactic compact binaries, and so on. 
TianQin will consist of three Earth orbiting satellites on nearly identical orbits with orbital radii of about $10^5$ km.
The satellites will form a normal triangle constellation whose plane is nearly perpendicular to the ecliptic plane. 
The TianQin project has been progressing smoothly following the ``0123" technology roadmap. 
In step ``0", the TianQin laser ranging station has been constructed and it has successfully ranged to all the five retro-reflectors on the Moon. 
In step ``1", the drag-free control technology has been tested and demonstrated using the TianQin-1 satellite. 
In step ``2", the inter-satellite laser interferometry technology will be tested using the pair of TianQin-2 satellites.
The TianQin-2 mission has been officially approved and the satellites will be launched around 2026. 
In step ``3", i.e., the TianQin-3 mission, three identical satellites will be launched around 2035 to form the space-based gravitational wave detector, TianQin, and to start gravitational wave detection in space.
\end{abstract}

%
\vspace{2pc}
\maketitle
%
%

\acrodef{GR}{General relativity}
\acrodef{GW}{gravitational wave}
\acrodef{EM}{electromagnetic}
\acrodef{QNM}{quasi normal mode}
\acrodef{MBHB}{massive black hole binary}
\acrodef{EMRI}{Extreme Mass Ratio Inspiral}
\acrodef{IMRI}{Intermediate Mass Ratio Inspiral}
\acrodef{SBHB}{stellar mass black hole binary}
\acrodef{GCB}{Galactic compact binary}
\acrodef{SGWB}{stochastic GW background}
\acrodef{VB}{verification binary}
\acrodef{IMBH}{intermediate-mass black hole}
\acrodef{MGT}{modified gravitation theory}
\acrodef{PN}{post-Newtonian}
\acrodef{ppE}{parameterized post-Einsteinian}
\acrodef{SNR}{signal-to-noise ratio}
\acrodef{AGN}{active galactic nuclei}
\acrodef{LIGO}{Laser Interferometer Gravitational-Wave Observatory}
\acrodef{CPL}{Chevallier-Polarski-Linder}
\acrodef{TM}{test mass}
\acrodef{TDI}{time delay interferometry}
\acrodef{TTL}{tilt-to-length}
\acrodef{GNSS}{Global Navigation Satellite System}
\acrodef{DFC}{drag-free control}
\acrodef{DFACS}{Drag-Free Attitude Control System}
\acrodef{S/C}{spacecraft}
\acrodef{FSU}{frequency stabilization unit}
\acrodef{ADC}{analog-to-digital converter}
\acrodef{OPL}{optical path length}
%

\tableofcontents
%
%
\section{Introduction}

Most effectively generated in highly dynamical processes involving massive and ultra-compact astrophysical objects, \acp{GW} are unlocking a path to observe the hidden sectors of the unverse with rich dynamical information.
The publication of the first credible \ac{GW} detection by the LIGO Scientific and Virgo Collaborations signifies the beginning of the \ac{GW} astronomy Era \cite{LIGOScientific:2016aoc}.
Efforts are now being made to build more powerful detectors and to open up more \ac{GW} spectrum for detection.
In the high frequency end ($10\sim10^4$ Hz), one may expect to see Cosmic Explorer \cite{Evans:2021gyd} and Einstein Telescope \cite{Maggiore:2019uih,Branchesi:2023mws} detecting the merger of stellar mass compact binaries to high redshifts ($z>2$) and at rates of one per every few minutes starting from the 2030s.
In the nanohertz frequency range (around $10^{-9}$ Hz), indications of a possible positive detection have been reported \cite{Xu:2023wog,EPTA:2023fyk,Reardon:2023gzh,NANOGrav:2023gor,IPTA:2023ero}.
At the lower end of the \ac{GW} spectrum, efforts are being made to look for possible signatures of primordial \acp{GW} through CMB observations (see, e.g., \cite{BICEP:2021xfz,Li:2017drr}).
Probably the most awaited is that, in the millihertz frequency range ($10^{-4}\sim1$ Hz), one may expect to see multiple space-based \ac{GW} detectors, such as LISA \cite{LISA:2017pwj}, TianQin \cite{TianQin:2015yph,TianQin:2020hid} and Taiji \cite{Hu:2017mde}, starting to operate around 2035 \cite{Gong:2021gvw}.
The Japanese DECIGO mission aims to detect \acp{GW} in the deci-Hertz frequency range \cite{Kawamura:2020pcg}.

Expected to be launched around 2035, the space-based \ac{GW} detector TianQin aims to detect \acp{GW} in the frequency range $10^{-4}\sim$ 1 Hz \cite{TianQin:2015yph}.
TianQin will be an equilateral triangle constellation consists of three drag-free satellites, orbiting the Earth with orbital radii of about $10^5$ km.
The detector plane of TianQin is nearly perpendicular to the ecliptic plane and the Sun will pass through the fixed orbital plane of TianQin every half year.
So TianQin adopts a consecutive ``three-month on + three-month off" detection scheme to protect the sensitive instruments from the direct illumination by the Sun.
TianQin aims to detect a variety of astrophysical and cosmological
\ac{GW} sources, populating different epochs of the universe, and is expected to boost astrophysics, fundamental physics and cosmology into completely new fronts with observation data that is never seen before.

\begin{figure}[ht!]
\centering
\includegraphics[width=\linewidth]{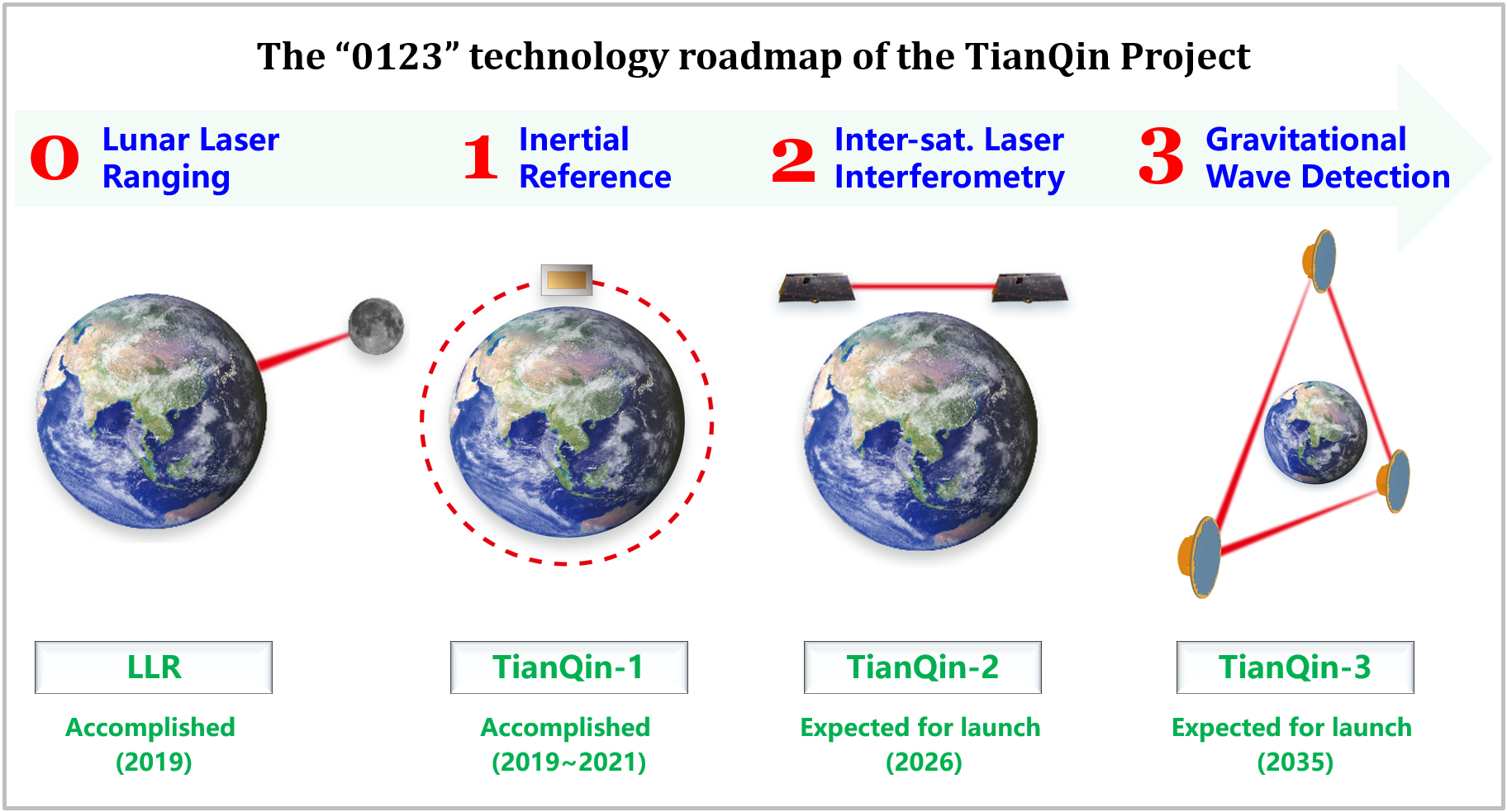}
\caption{The ``0123" technology roadmap of the TianQin Project. See main text for more explanation.}
\label{fig:roadmap}
\end{figure}

The ``0123" technology roadmap (Fig. \ref{fig:roadmap}) has been used to guide the development of the TianQin project:
\begin{itemize}
\item Step ``0": To acquire the capability of lunar laser ranging, with which one can obtain high-precision orbit information for the TianQin \ac{S/C};
\item Step ``1": To use a single satellite to test and demonstrate the inertial reference technology;
\item Step ``2": To use a pair of satellites to test and demonstrate the inter-satellite laser interferometry technology;
\item Step ``3": To launch a constellation of three satellites to form the space-based \ac{GW} detector, TianQin, and to carry out \ac{GW} detection in space.
\end{itemize}

Both step ``0" and step ``1" have been successfully carried out, which have prepared TianQin to precisely range to the TianQin satellites with lasers and have demonstrated the principle capability of \ac{DFC} for TianQin. Step ``2" has been officially approved in 2021 and a pair of TianQin-2 satellites are expected to be launched around 2026 to demonstrate the inter-satellite laser interferometry technology for TianQin. Various preparation work for the step ``3", including science study, mission study and technology development, has also been carried out.

In this paper we present an update on the various aspects of the TianQin project.

\section{Science Objectives of TianQin}

{\it Coordinator: Jianwei Mei}

By detecting \acp{GW} in the milli-hertz (0.1 mHz $\sim$ 1 Hz) range, TianQin is sensitive to a variety of \ac{GW} sources, including:

\begin{itemize}
\item The revolution of \ac{GCB};
\item The inspiral of \ac{SBHB};
\item The \ac{EMRI} and \ac{IMRI};
\item The inspiral-merger-ringdown of \ac{MBHB};
\item The \ac{SGWB}, including those from unresolved astrophysical sources and from possible energetic processes in the early universe.
\end{itemize}

There are notable features about the types of \ac{GW} signals expected for TianQin.
Firstly, TianQin can detect \ac{GW} sources from different epochs of the universe, for example:
\ac{GCB} systems in the Galaxy \cite{Huang:2020rjf},
\ac{SBHB} systems to redshifts of the order $z\sim\cO(0.1)$ \cite{Liu:2020eko},
\ac{EMRI} systems to redshifts of the order $z\sim\cO(3)$ when the star formation rate was at its peek \cite{Fan:2020zhy},
\ac{MBHB} systems to redshifts of the order $z\sim\cO(10)$ when the first stars and galaxies just appeared \cite{Wang:2019ryf},
and possibly also first order electroweak phase transitions when the universe was only about $\cO(10^{-10}$ s) old \cite{Liang:2021bde}.
Secondly, TianQin can detect some \ac{GW} signals with extremely high \acp{SNR}. For example, \acp{SNR} for some \acp{MBHB} signals can reach the order of $\cO(10^3)$ \cite{Wang:2019ryf}.
Thirdly, TianQin can measure the source parameters of some \ac{GW} signals to extremely high precision. For example, some of the parameters of \acp{SBHB}, \acp{EMRI} and \acp{MBHB} can be measure to better in $\cO(10^{-6})$ \cite{Liu:2020eko,Fan:2020zhy,Wang:2019ryf}.

These capabilities promise a huge scientific discovery space for TianQin alone \cite{TianQin:2020hid,Li:2024rnk,Luo:2025tqr} and in joint detection with other space-based detectors such as LISA \cite{Torres-Orjuela:2023hfd}. Potential scientific discoveries include:
discovering massive and \ac{IMBH} binary systems and compact binary systems in globular clusters;
revealing the structure and material distribution of the Milky Way, the central environment of galaxies surrounding massive black holes, and the formation and evolution mechanisms of compact binary systems;
high precision test of general relativity in the strong field regime;
accurate characterization of the evolution history of the universe, and precise measurement of dark matter and dark energy related parameters in a wide span of the cosmic history;
and revealing the dynamics of symmetry breaking and first-order phase transition in the early universe.

\subsection{Astrophysics with TianQin}

{\it Coordinator: Yi-Ming Hu}

TianQin will have the ability to detect sources across a wide mass spectrum, over a wide range of cosmic history.
The successful detection of the mergers of compact objects like white dwarfs, neutron stars, and massive black holes, can reveal rich information on the birth and growth of stars, galaxies and black holes.
Coupled with unparalleled observation precision in parameters like mass and spin, TianQin can shed light on long-standing puzzles like the formation channel of stellar-origin black holes, seeding mechanism of massive black holes, and the stellar environments around massive black holes.

In this subsection, we aim to provide a brief overview on the detection ability of TianQin on the major types of \ac{GW} sources.
We discuss how the future detections, either by TianQin alone, or in coordination with other equipments, can re-shape our understanding on stellar physics, the birth and growth of massive black holes, and the immediate environment around massive black holes.
A more thorough discussion can be found in \cite{Li:2024rnk}.

\subsubsection{Detection ability of TianQin}\

Hundreds of millions of \ac{GCB} are expected to exist in our Galaxy.
Even if only a tiny fraction of them are observed by TianQin, the total detection number can still reach the level of tens of thousands.
Their observation and identification with the electromagnetic telescopes are challenging and only about a dozen of the known \acp{GCB} are detectable by TianQin.
These are known as the \acp{VB} \cite{Stroeer:2006rx}.
Some \acp{VB} can have very high \ac{SNR}.
For example, the nominal reference source for TianQin, RX J0806.3+1527 (also known as HM Cancri, here after J0806), can accumulate \ac{SNR} of 5 within two days, and exceed 100 after the five year mission lifetime.
TianQin can also measure the physical parameters precisely.
For examples, most of the \acp{GCB} can be localized to $\cO(1{\rm~deg}^2)$, the frequency parameters can be constrained to the level of $\cO(10^{-7})$, and the amplitude to the $20\%$ level \cite{Huang:2020rjf}.

The other promising source for TianQin is the early inspiral of \ac{SBHB}.
Ground-based \ac{GW} detectors have made hundreds of detections of \ac{SBHB} mergers over the course of nearly a decade of operation and upgrade.
Assuming the observed population of \acp{SBHB}, combined with TianQin's sensitivity curve, it can be deduced that a handful of \acp{SBHB} can be detected by TianQin, and the detection number can be increased either by lowering the detection threshold or by collaborating with other facilities.
Many source parameters can be constrained very precisely.
For examples,
the merger time can be measured to $\cO(1{\rm~s})$,
the location of the source can be determined to $\cO(0.1$ deg$^2$),
and the masses can be constrained to the $\cO(10^{-6})$ level \cite{Liu:2020eko,Liu:2021yoy}.

TianQin is expected to detect \ac{EMRI} signals but the detection rate is subject to large theoretical uncertainties.
Except for a few most pessimistic models, most theoretical models predict that TianQin can detect at least a couple of \ac{EMRI} signals, with the most optimistic ones predicting hundreds of detections per year.
For the detected events, all intrinsic parameters can be precisely constrained to the level of $\cO(10^{-6})$, while the extrinsic parameters can be constrained to the 10\% level \cite{Fan:2020zhy}.
TianQin also has the potential to detect the extreme mass-ratio bursts when the \acp{EMRI} are just forming \cite{Fan:2022wio}.

There is also large theoretical uncertainty on the number of \ac{MBHB} merger events that TianQin can detect.
But if there is a detectable source, TianQin can detect it to the edge of the observable Universe.
Even for \acp{MBHB} merging at very high redshift (such as $z \sim 15$), TianQin can still make clear detections and precise measurements of their distance and mass parameters.
Apart from help revealing the origin and growth history of massive black holes, such capability can also be invaluable for exploring other aspects of the Universe, such as the global structure of the cosmic space \cite{Shi:2024ula}.
For nearby \ac{MBHB} merger events, TianQin can pinpoint the location and issue early warnings at time scales such as a day before merger \cite{Wang:2019ryf}.

\ac{SGWB} contains the information of unresolvable \ac{GW} events.
The most obvious component is the foreground from the \acp{GCB}.
With a five-year mission of TianQin, the \ac{SNR} of the \ac{SGWB} foreground can exceed 100 \cite{Liang:2021bde}.

\subsubsection{Stellar physics}\

Through detailed analysis of stellar mass compact objects, TianQin can help deepening our understanding of binary evolutions.

In the lower end of the mass spectrum, TianQin's observation of \acp{GCB} can assess the mass of compact systems, unveil the nature of the mass gap (about $2.5\sim5\msun$) between the lightest black hole and the heaviest neutron stars.
The large number of detections of \acp{GCB} provide a priceless catalog that can help assess the Galactic structure, tidal physics, SNe Ia progenitors, etc.
In combination with electromagnetic detectors, TianQin can also increase our understanding of ultra-compact X-ray binaries (UCXBs) and AM CVns \cite{Wang:2023joh,Kupfer:2023nqx,Szekerczes:2023gct,Shao:2021dbg}.

Standard stellar evolution models predict no black holes in the mass gap at about $65\msun\sim120\msun$ due to pair-instability supernova.
However, \ac{GW}190521 challenged this.
TianQin and LISA can detect such systems in the early inspiral phase when the orbital eccentricities still carry the imprint of formation channels.
By measuring these eccentricities, we can distinguish between different formation scenarios, such as those from isolated binaries with very low metallicity and hierarchical mergers between black holes/stars.
Notice that the good sensitivity of TianQin in high frequency leads to an advantage in detecting \acp{SBHB} beyond the pair-instability mass gap \cite{Liu:2021yoy,Toubiana:2020drf,Sberna:2022qbn}.

In dense star environments like Pop III clusters, black holes can form hierarchical triple systems. The orbital evolution of inner binary within merging triples is different from that of isolated binaries due to the perturbation from the third object. TianQin can distinguish their formation channels by measuring the orbital eccentricities of the merging \acp{SBHB} \cite{Wang:2023tle}.
Considering a binary moving around an massive black hole, gravitational perturbation from the massive black hole causes the inner binary to precess and may induce eccentricity oscillations.
\ac{GR} effects involving the massive black hole can generate extra precessions and change the dynamics.
TianQin can observe these effects and depict the three-body dynamics \cite{Sakurai:2017opi,Zwick:2024yzh,Liu:2021uam}.

The mergers of stellar-mass black holes could also happen in \ac{AGN} disks. Compact objects in \ac{AGN} disks can form binaries through dynamical interactions and in-situ star formation. TianQin can reveal the evolutionary processes of binaries, constrain accretion rates and binary masses, and determine the contribution of \ac{AGN} channel. It can also detect the Doppler acceleration and shift of mergers near massive black holes, unveil the locations of gaps and migration traps, and help identify potential associations of electromagnetic emissions with \acp{GW} \cite{Secunda:2018kar,Secunda:2020mhd,Samsing:2020tda}.

\subsubsection{Birth and growth of massive black holes}\

TianQin can help distinguish between different massive black hole seeding mechanisms.
The two main models are the light-seed model, where massive black holes originate from the Pop~III star remnants, and the heavy-seed model, involving the direct collapse of gas clouds.
By detecting \acp{GW} from merging binary black holes across a wide range of masses, TianQin can test the predictions of different seeding models.
For example, the initial mass function of seed black holes can be investigated.
If Pop~III stars are the main seeds, then their remnants would have relatively lower masses compared to those from direct collapse.
TianQin's ability to detect lower-mass sources in the millihertz frequency range makes it well-suited for this task \cite{Ding:2019tjk,Zwick:2022mcu}.

In colliding galaxies, \acp{IMBH} can form binaries when the clusters containing them merge. TianQin's sensitivity to higher frequencies allows it to observe these events. By studying the dynamical evolution of \ac{IMBH} binaries, such as their eccentricities and inclinations, we can learn about the dense stellar environments in which they form \cite{Torres-Orjuela:2023hfd}.

When massive black holes merge, there may be accompanying electromagnetic signals.
TianQin's real-time data transmission capability is crucial for enabling multi-messenger observation. By quickly identifying and localizing the source, it enables follow-up observations with electromagnetic telescopes. The pre-merger stage may show quasi-periodic variations in disk luminosity, while the post-merger stage can have changes in disk and jet luminosity due to the recoil of the new black hole.
In \ac{AGN} environments, \ac{IMBH} binaries can produce associated flares when they merge. TianQin's ability to work with other observatories in a multi-messenger approach can enhance our understanding of these processes \cite{Chen:2023qga}.

\subsubsection{Environment surrounding massive black holes}\

TianQin's observation of \ac{EMRI} and \ac{IMRI} signals can reveal secrets about the sources.

Firstly, it provides insights into the existence and properties of \acp{IMBH}.
By detecting light \acp{IMRI}, which consist of an \ac{IMBH} and a stellar-mass black hole, TianQin can confirm the presence of \acp{IMBH} with masses in the range of $10^{2}-10^{5}\,{\rm M_\odot}$. The formation of \acp{IMBH} can occur through various pathways such as runaway mergers of massive stars in dense star clusters or repeated mergers of black holes in binary systems in a dense dynamical environment. Through the detection of \acp{IMRI}, one can better understand the growth processes of these black holes and their interactions within the host systems \cite{Amaro-Seoane:2018gbb,Arca-Sedda:2020lso,Torres-Orjuela:2023hfd}.

Secondly, TianQin can help understand the environmental effects on compact binary systems.
In dense stellar systems like nuclear star clusters or globular clusters where \acp{IMRI} are expected to form, there are effects like peculiar velocity and Brownian motion.
TianQin can detect the changes in the \ac{GW}  modes due to the peculiar velocity of the host system, which helps in determining the velocity of the source.
For example, for \acp{IMRI} with different total masses, the measurability of the velocity lies in the detection of changes in non-quadrupolar modes.
Also, one can detect the acceleration of \acp{IMRI} caused by Brownian motion by observing the phase shift in the \acp{GW}.
This provides information about the dynamics of the star clusters and the interaction of \acp{IMRI} with the surrounding stars \cite{Torres-Orjuela:2020dhw,Torres-Orjuela:2020oxq}.
In addition, when \acp{IMRI} are in the accretion disks of \acp{AGN}, TianQin can detect the gas-induced dephasing effect.
By calculating the orbital evolution and phase shift of \acp{IMRI} in the gas environment, we can study the interaction between the binary system and the gas.
This is important for understanding the evolution of \acp{IMRI} in the \ac{AGN} disks and the impact of the gas environment on the \ac{GW} emission \cite{Yunes:2011ws,Barausse:2014tra,Garg:2022nko}.

Finally, in multi-messenger astronomy, TianQin's detection of \ac{GW} sources accompanied by electromagnetic radiation helps in identifying the host systems and the understanding of the astrophysical processes \cite{Li:2024rnk,Torres-Orjuela:2023hfd}.
For example, in the case of \acp{EMRI} and \acp{IMRI} in \acp{AGN} with a gas environment or with a star as the minor component, there can be electromagnetic radiation during or after the merger.
TianQin can detect the \acp{GW} and, combined with time-domain surveys, help in identifying the host \ac{AGN}. For tidal disruption events, TianQin can detect the \acp{GW} produced when a star is tidally disrupted by a massive black hole, and together with the observed electromagnetic emissions, this can help understand the disruption process and constrain the mass function of the massive black holes \cite{Pfister:2021ton,Ye:2023aeo}.

\subsection{Fundamental Physics with TianQin}\label{sec:sci_phy}

{\it Coordinator: Jian-dong Zhang}

As a theory of gravity, \ac{GR} has passed numerous experimental tests.
However, most of the tests are performed in the weak field regime \cite{Will:2014kxa}.
With the breakthrough in \ac{GW} detection, the \ac{GW} signals of binaries are used in the study of fundamental physics \cite{LIGOScientific:2016lio,LIGOScientific:2019fpa,LIGOScientific:2020tif,LIGOScientific:2021sio}.
The most remarkable fact about \ac{GW} is that it can test the fundamental physics in the strong field regime.
As a space-based \ac{GW} detector, TianQin could detect some \ac{GW} signals with very high \ac{SNR} and precision, thus can push the test of \ac{GR} to complete new fronts.

In this subsection, we summarize how TianQin can help verify the key predictions of \ac{GR} in the strong field regime and search for possible signatures of beyond \ac{GR} effects.
We also discuss the environmental effects that may interfere with such effort.
A more thorough discussion of the results can be found in \cite{Luo:2025tqr}.

\subsubsection{Testing the key predictions of GR}\

The experimental test of \ac{GR} has been going on for over a century \cite{Will:2014kxa}.
The detection of \ac{GW} has been another remarkable success for \ac{GR} \cite{LIGOScientific:2016lio}.
But only with space-based \ac{GW} detection that some of the key predictions of \ac{GR} can be robustly tested for the first time.

\paragraph{Higher modes and nonlinear modes}

Nonlinearity is a characteristic feature of Einstein's equations, which could be studied with \acp{GW}.
The \ac{GW} signal from a collapsing binary typically includes three stages: inspiral, merger and ringdown. For the early inspiral and ringdown stages, one can use the perturbation methods to obtain the waveforms.

For example, the ringdown signal after the merger of a \ac{MBHB} can be expanded in terms of a series of \acp{QNM} \cite{Kokkotas:1999bd,Berti:2009kk,Konoplya:2011qq}.
So far only \acp{GW} from the linear order have been confirmed in ground-based detectors.
Among the linear modes, the $(220)$-mode, which has $\ell=m=2$ and $n=0$, is the fundamental mode, and all other modes are called higher modes and overtones.
Limited by the sensitivity of existing ground-based detectors, most of the detected \ac{GW} events have \acp{SNR} of 30 or less \cite{LIGOScientific:2021sio}, while the \acp{SNR} of the ringdown phase are even weaker.
Thus no higher modes has been confirmed in the existing \ac{GW} data.

By using the numerically fit amplitude formulae \cite{London:2014cma}, the prospect of using TianQin to detect 11 different higher modes and nonlinear modes has been studied in \cite{Shi:2024ttu}.
It has been found that some of the \acp{SNR} can reach a few dozens even for the source redshift $z=3$.
After considering three different astrophysical models, the detection numbers for each of the 11 higher modes and nonlinear modes have been analysed.
Apart from the (4,3,0) mode, all other modes are expected to be detected in at least in one \ac{MBHB} event.

\paragraph{Memory effect}

The radiation of \acp{GW} will result in a permanent change to the background spacetime.
Such change is related to the entire history of \ac{GW} radiation, and the phenomenon is referred to as the \ac{GW} memory effect \cite{Braginsky:1987kwo}.
The memory effect is one of the direct predictions of \ac{GR} in the nonlinear and strong-field regime, and so the detection of the memory effect is a direct test of \ac{GR}.

The prospect of using TianQin to detect the memory effect has been studied in \cite{Sun:2022pvh,Sun:2024nut}.
Based on a few astrophysical population models of \acp{MBHB}, it has been found that TianQin can detect approximately 0.5 to 2 \ac{MBHB} merger events for which the displacement memory effect can have \acp{SNR} greater than 3.
The chance for TianQin to detect the spin memory effect from a single \ac{MBHB} event is found to be negligible  \cite{Sun:2022pvh}.
The memory effect can help break the degeneracy between the inclination angle and luminosity distance during parameter inference.
By calculating the Bayes factor, it has been found that an \ac{SNR} of approximately 2.36 is sufficient for TianQin to claim the detection of the memory effect \cite{Sun:2024nut}.

\paragraph{Kerr hypothesis}

The no-hair theorem claims that the black holes in \ac{GR} can be fully characterized by mass, spin, and electric charge.
However, astrophysical black holes are believed to be nearly neutral due to several charge loss and neutralization mechanisms \cite{Gibbons:1975kk,Goldreich:1969sb,Ruderman:1975ju,Blandford:1977ds}.
Thus it's believed that the black holes in our universe can be described by the stationary and rotating Kerr metric \cite{Kerr:1963ud}.
This is called the Kerr hypothesis.
With \acp{GW}, high-precision tests of the Kerr hypothesis can be conducted through different ways, such as detecting the ringdown signal of a black hole \cite{Dreyer:2003bv} and measuring the multipole moment of a black hole \cite{Ryan:1995wh}.

If \ac{GR} and the Kerr hypothesis is valid, the oscillation frequencies and damping times of the \acp{QNM} are entirely determined by the mass and spin of the final Kerr black hole.
One can test the Kerr hypothesis by measuring multiple frequencies and damping times of the \acp{QNM}, and check if they correspond to the same mass and spin.
The oscillation frequency $\omega_{lmn}$ and the damping time $\tau_{lmn}$ can be parameterized as:
\be
	\omega_{lmn}=\omega_{lmn}^{GR}(1+\delta\omega_{lmn}),~~~\tau_{lmn}=\tau_{lmn}^{GR}(1+\delta\tau_{lmn})
\ee
where the deviation parameters $\delta\omega_{lmn}=\delta\tau_{lmn}=0\,$ if \ac{GR} is correct.
For TianQin, the combination of $\delta \tau_{22}$, $\delta\omega_{22}$, and $\delta w_{33}$ offers the most stringent constraint for vast majority of cases.
With the consideration of three astrophyiscal models, the selected deviation parameters can always be constrained to the 1\% level or better, and some can even reach the $\cO(10^{-4})$ level \cite{Shi:2019hqa}.

For an isolated massive object, its gravitational field can be characterized by its multipole moments.
The multipole moments of a Kerr black hole are fully determined by its mass $M$ and spin $a$:
\bea	
M_l+iS_l=M(ia)^l\,,\quad l=0,1,2,\cdots\,.
\label{eq:moment}
\eea
To test if an astrophysical black hole is a Kerr black hole, one can measure the quadrupole moment with $l=2$, in addition to the mass and spin.
One can parameterize such test by treating the quadrupole moment as an additional parameter, $Q=-(1+\delta\kappa) a^2 M$.
$\delta\kappa$ is a deviation parameter depending on the internal structure of the object, and it equals to $0$ for Kerr black holes.
The prospect of using TianQin to measure the Kerr quadrupole moment has been studied in \cite{Zi:2021pdp,Kong:2024ssa}.
Using \acp{MBHB}, TianQin can constrain $\delta\kappa$ to the order $\cO(10^{-2})$, and the events with asymmetric mass will have better capability.
It has been found that TianQin can constrain $\delta\kappa$ to the order $\cO(10^{-6})$ using \acp{EMRI}.
Comparing to the mass, the spin of the central black hole has a more significant impact on the constraints, and the larger the spin, the stronger the constraints.

\subsubsection{Looking for possible signatures of beyond GR effect}\

If \ac{GR} is not fully correct, one may hope to find some beyond \ac{GR} effect in \ac{GW} emission, including the fundamental degrees of freedom of \acp{GW}, the generation and propagation of \acp{GW}.
The possible beyond \ac{GR} effects can be searched with various parameterization schemes, with various \acp{MGT} being representative examples.
Current studies have not found any deviations from \ac{GR}, but TianQin can push the tests to new frontiers.

\paragraph{GW polarization}

In \ac{GR}, \acp{GW} possess only two tensor polarization modes.
But for a general metric theory of gravity, the metric tensor has 6 propagation degrees of freedom,
thus there can exist 6 polarization modes \cite{Eardley:1973br,Eardley:1973zuo}.
These additional polarization modes can be excited by the coupling between the metric and extra gravitational fields.

For the \ac{GW} events detected by ground-based detectors, a Bayesian model selection analysis shows that the data supports the assumption that the signals are consisted of purely tensor modes, rather than purely vector or scalar modes \cite{LIGOScientific:2017ycc,LIGOScientific:2019fpa,LIGOScientific:2018dkp}.
The null-stream  method has been used for the O2, O3a and O3b events, and all the data is consistent with the pure tensor mode hypothesis \cite{Pang:2020pfz,Wong:2021cmp}.

TianQin is expected to detect about $10^4$ pairs of \acp{GCB} \cite{Huang:2020rjf} and the prospect of using TianQin and \ac{GCB} signals to search for extra polarization modes has been studied in \cite{Xie:2022wkx}.
Due to the vanishing of the antenna pattern function, TianQin has no detection power for the vector or scalar modes for sources located in the direction of J0806 and its antipodal point.
For sources located in other directions, the best precision on $\alpha_v$ can reach the 2\% level and that for $\alpha_s$ can reach the 5\% level, where $\alpha_v$ and $\alpha_s$ are the relative magnitudes of the vector and scalar modes compared to the two \ac{GR} modes, respectively.
\acp{VB} with known position can produce better constraints, and ZTF J1539 is currently the best among all the \acp{VB}.

The \ac{MBHB} signals can also be used to constrain the extra polarization modes, and the correction on the phase evolution must be considered in the waveform.
In a preliminary study with the Bayes method, the constraint with TianQin on the amplitudes of extra polarization modes has been found to be about a few percent \cite{Ning:2024xxx}.
The possibility of using \ac{SGWB} to constrain the extra polarization modes has been studied in \cite{Hu:2023nfv,Hu:2024toa}.

\paragraph{GW propagation}

In \ac{GR}, \acp{GW} travel at the speed of light.
But \acp{GW} in \acp{MGT} may propagate differently than the speed of light.
In general, we can consider a non-trivial dispersion relation for \acp{GW} as
\be
E^2=p^2+\mathbb{A}_\alpha p^\alpha\,,
\ee
where $E$ and $p$ are the energy and momentum of the graviton, respectively.
$\alpha$ is a power and $\mathbb{A}_\alpha$ is the corresponding magnitude of modification.
In \ac{GR}, $\mathbb{A}_\alpha=0$.
For $\alpha=0$, one usually writes $\mathbb{A}_0=m_g^2$, where $m_g$ corresponds to the graviton mass.
The analysis of the GWTC-1 \cite{LIGOScientific:2018mvr}, GWTC-2 \cite{LIGOScientific:2020ibl} and GWTC-3 \cite{KAGRA:2021vkt} data has set the bound to $m_g\leq 1.27\times10^{-23}$ eV.
The current bound on $\mathbb{A}_\alpha$ for $\alpha \in [0,4]$ can be found in \cite{LIGOScientific:2019fpa,LIGOScientific:2020tif,LIGOScientific:2021sio}.

A preliminary study shows that TianQin can probe the graviton mass to $m_g<\cO(10^{-27}\rm~eV)$, thus improving over the current bound on graviton mass by four orders.
TianQin can also improve over the current bounds on $\mathbb{A}_0$, $\mathbb{A}_{0.5}$ and $\mathbb{A}_1$ by about eight, five and three orders, respectively \cite{Luo:2025tqr}.

\paragraph{GW generation}\label{subsec:ppE}

The \ac{ppE} framework has been developed to enable a theory agnostic probe of possible deviations from \ac{GR} \cite{Berti:2004bd,Arun:2006yw,Arun:2006hn,Yunes:2009ke}.
The basic idea of \ac{ppE} is to focus on the leading \ac{PN}-order corrections to the \ac{PN} waveform,
\bea
h_{\rm ppE}(f)=h_{\rm {GR}}(f)(1+\alpha u^a)e^{i\beta u^b}\,,\label{eq:waveform_ppE}
\eea
where $\alpha$ and $\beta$ are the \ac{ppE} parameters, with $\alpha=\beta=0\,$ in \ac{GR}, and $a$ and $b$ are the \ac{PN}-order parameters, with $b=k-5\,$ and $a=b+5$ corresponding to the $(k/2)$ \ac{PN} order.
The \ac{ppE} parameters for a given \ac{MGT} can be found by computing the corrections to the orbital evolution of the binary system \cite{Tahura:2018zuq}.
Using this approach, \ac{ppE} parameters have been determined for a spectrum of theoretical models,
See \cite{Tahura:2018zuq,Chamberlain:2017fjl} for a summary.

The prospect of using TianQin to test \acp{MGT} with the \ac{ppE} formalism has been studied in \cite{Shi:2022qno}.
The result shows that $\Delta\beta$ is more tightly constrained by the low-mass sources at the lower \ac{PN} orders, while $\Delta\beta$ at the higher \ac{PN} orders are best constrained with the sources at around $\cO(10^5\msun)$.
The prospect of testing specific \acp{MGT} has also been studied.
With the detection of \acp{SBHB} with total masses below $M<\cO(10^2\msun)$, TianQin is expected to constrain EdGB to the order $\sqrt{|\bar{\alpha}_{\rm EdGB}|}<\cO(0.1{\rm~km})$, which is about one order better than the current best result.
Similarly, TianQin is expected to constrain dCS to the order $\sqrt{\bar{\alpha}_{\rm dCS}}<\cO(1{\rm~km})$.
In the case of non-commutative gravity, TianQin is expect to improve over the current best bound by an order of magnitude and constrain the theory to the sub-Planckian scale.
For $\dot{G}$, TianQin can push the constraint to the level of $|{\dot{G}/G_0}|<\cO(10^{-5}\rm~year^{-1})$.
For more details, we refer to \cite{Shi:2022qno}.

\subsubsection{Environmental effects}\

In searching for possible signatures of beyond \ac{GR} effect, an important issue is to avoid mistaking false signals for evidence of new physics.
Around the \ac{GW} radiating binary sources, there may exist accretion disks, dark matter halos, or third gravitational bodies.
The surrounding matter can change the orbit evolution due to the gravitational pull and dynamical friction, or change the mass and spin of the sources due to the accretion of surrounding matter.
On the path of the \ac{GW} propagation, there can exist different density of matter, causing gravitational lensing effect for \acp{GW}.
Depending on the density profile of the lenses, \ac{GW} can be bent, delayed, (de)magnified, phase-shifted, and diffracted.
In real \ac{GW} detection, the environment is not known.
So the problem is how to distinguish between the environmental effect and possible signatures of beyond \ac{GR} effect.

\paragraph{Environmental effect in \ac{GW} generation}

The capability of TianQin in probing the environmental effect during \ac{GW} generation can be directly obtained from the \ac{ppE} result (see subsection \ref{subsec:ppE}).
For example, the dynamical friction due to a dark matter spike with density profile $\rho_{\rm DM}=\rho_0(r_0/r)^{3/2}$ affects the inspiral signal at the $-4$ \ac{PN} order, which is the same as the effect of $\dot{G}$.
To distinguish these two effects, one can define the following function,
\bea F=\sum_{i=1}^n\frac{(\dot G_i-\bar{\dot {G}})^2}{\sigma_i^2}\,,\eea
where $n$ is the number of detected events and the index $i$ means the $i$-th event,
$\dot G_i$ and $\sigma_i$ are the mean value and variance of $\dot G$ for the $i$-th event.
$\bar{\dot {G}}$ is the mean value of $\dot G_i$ for all the events.
By using a specific astrophysical population model for binary black holes, one can find that $F$ will be small if the waveform correction is due to the varying-G, while $F$ is very large if the correction is due to dark matter halo \cite{Yuan:2024duo}.
This result shows that it is possible to distinguish these two effects if multiple events are detected.

\paragraph{Environmental effect in \ac{GW} propagation}

When electromagnetic waves pass by a massive object, there will be gravitational lensing effect.
Similar to electromagnetic waves, \acp{GW} can also be lensed \cite{Takahashi:2003ix}.
If the lensing effect is not properly included in the analysis of \ac{GW} data, there can be systematic errors in the estimation of source parameters \cite{Liu:2024xxn}.
What's more, lensed \ac{GW} signals can be used to study the propagation property of \acp{GW}, infer the physical properties of the lensing object, study the nature of dark matter and the expansion of the universe.

So far the LIGO-Virgo-KAGRA collaboration has published 90 \ac{GW} events \cite{LIGOScientific:2021aug}.
Despite much effort, however, no lensed \ac{GW} signal has been confirmed in these events \cite{LIGOScientific:2023bwz}.
Recent study suggest that nearly one percent of the detected events for TianQin may experience strong gravitational lensing  \cite{Gao:2021sxw}.
It is also possible for wave-optics effects of lensing to be detected \cite{Caliskan:2022hbu,Tambalo:2022wlm}, if the \ac{GW} wavelength is comparable or longer than the gravitational radius of the lens.

The prospect of using TianQin to probe the gravitational lensing effect for \acp{GW} has been studied in \cite{Lin:2023ccz}.
The result shows that the gravitational lensing increases both the \ac{SNR} and the precision of parameter estimation.
The parameter of the lens can also be measured with an accuracy of $\cO(10^{-5})$ or better.

\subsubsection{New fundamental matter and interactions}\

{\it Coordinator: Fa Peng Huang}

In this part we discuss the prospect of using TianQin to probe possible new physics in the non-gravitational sector.
To explain the origin of matter-antimatter asymmetry in the observable universe and the microscopic nature of dark matter, it is often necessary to introduce new fundamental particles and interactions beyond the Standard Model of particle physics.
Over the past few decades, experimentalists have not observed these new particles or interactions in dark matter direct detection experiments, collider experiments, or other related studies. This may suggest the need for experimentalists to explore new experimental approaches.
As a space-based \ac{GW} detector, TianQin could open a unique and novel window to probe such new fundamental matter and interactions.

\paragraph{Probing the nature of dark matter}

Dark matter may directly generate \ac{GW} signals during its production in the early universe or leave significant imprints on \ac{GW} signals throughout its astronomical evolution. Based on the current state of experimental and theoretical research on dark matter detection, attention has shifted toward studying ultralight or superheavy dark matter. TianQin has the potential to detect these two types of dark matter candidates through \ac{GW} signals.

Boson clouds formed by dark matter particles like axions through superradiance around black holes can affect the orbital evolution of binary black holes and neutron star-black hole systems, altering the \acp{GW} of such events \cite{Zouros:1979iw,Detweiler:1980uk,Dolan:2007mj,Arvanitaki:2014wva, Zhang:2019eid,Xie:2022uvp}. These effects can be used to reveal dark matter properties near black holes through \ac{GW} detection \cite{Eda:2013gg,Eda:2014kra,Zhang:2019eid,Xie:2022uvp}.
Scalar and vector boson clouds can also emit \acp{GW} directly via pair annihilation of bound-state particles and energy-level transitions \cite{Arvanitaki:2010sy,Arvanitaki:2014wva,Arvanitaki:2016qwi}. The \ac{GW} signal from boson clouds is quasi-monochromatic, with frequency depending on the boson particle mass. These \acp{GW} can be detected individually by ground-based and space-based observatories \cite{Brito:2017wnc,Brito:2017zvb,Baryakhtar:2017ngi,Siemonsen:2019ebd,Palomba:2019vxe,LIGOScientific:2021rnv}, and they can also contribute to the stochastic background \cite{Brito:2017wnc,Brito:2017zvb,Tsukada:2020lgt,Yang:2023aak}. For example, the axion cloud's effect on binary \acp{GW} can be detected by TianQin.

In the early universe, the production of dark matter is often accompanied by the generation of \acp{GW}. Therefore, \ac{GW} experiments like TianQin can provide novel methods for probing the properties and production mechanisms of dark matter.  Different dark matter models predict different phase transition parameters, such as the phase transition strength \cite{Wang:2020jrd}, phase transition duration, and the bubble wall velocity \cite{Moore:1995si,Wang:2020zlf,Jiang:2022btc,Laurent:2022jrs}.
Phase transitions also offer new mechanisms for dark matter production, including filtered dark matter \cite{Baker:2019ndr,Chway:2019kft,Jiang:2023nkj},  soliton dark matter \cite{Krylov:2013qe,Huang:2017kzu,Hong:2020est,Jiang:2023qbm,Jiang:2024zrb}, and so on.

Interpreting \ac{GW} data to understand dark matter properties is complex, necessitating detailed modeling of both \ac{GW} sources and dark matter interactions. Additionally, distinguishing potential dark matter signals from other astrophysical sources requires highly sensitive and precise measurements.

\paragraph{Probing the origin of matter-antimatter asymmetry of the universe and the new Higgs potential}

After the discovery of the Higgs boson, exploring the shape of the Higgs potential has become a critical issue in particle cosmology. It also provides essential conditions for explaining the origin of matter-antimatter asymmetry in the universe.
A generic new physics model with new particles and new interactions would prediction new Higgs potential in the form of the dimension-6 operators \cite{Zhang:1992fs,Grojean:2004xa,Huang:2015izx,Huang:2016odd,Cai:2017tmh}. This new Higgs potential can induce  a strong first-order phase transition in the early universe and provide the necessary condition for the electroweak baryogenesis mechanism that naturally explains the origin of matter-antimatter asymmetry.
TianQin is expected to be capable of probing new physical models or new Higgs potential functions that can produce first-order electroweak phase transitions in the universe. Specifically, it can probe the parameter space of new physical models with a phase transition strength greater than $0.1$, namely, $\alpha>0.1$.

Various theoretical models have been proposed to explain matter-antimatter asymmetry through mechanisms involving \acp{GW}. Each model predicts specific \ac{GW} signals that can be tested by future observations. Mechanisms for achieving baryogenesis include electroweak baryogenesis, leptogenesis, and first-order phase transitions in the early universe. These often predict distinct \ac{GW} signatures detectable by various observatories. The \ac{GW} signals from phase transitions provide a direct probe of early universe conditions leading to baryogenesis, with different models predicting varying \ac{GW} spectra based on phase transition specifics.


\subsection{Cosmology with TianQin}   \label{sec:sci_cosmo}

{\it Coordinator: Liang-Gui Zhu}

After nearly a century of development, current cosmological research has entered the era of precision cosmology.
The standard model of cosmology --- the cosmological constant cold dark matter ($\Lambda$CDM) model,
describes well most of the cosmological observations from primordial nucleosynthesis to the present.
However, the $\Lambda$CDM model is not perfect. As the precision of measurements of various cosmological parameters improves,
the $\Lambda$CDM model faces several challenges,
the two most notable of which are:
(i) the inconsistencies in measurements of the Hubble-Lema\^itre constant ($H_0$) from different probes (commonly referred to as the Hubble tension) \cite{Planck:2018vyg, Riess:2021jrx, Riess:2024vfa, Freedman:2017yms, Riess:2019qba, Perivolaropoulos:2021jda, DiValentino:2021izs, Schoneberg:2021qvd, Cai:2021weh}
and (ii) the significant deviation of the dark energy equation of state from the cosmological constant \cite{Zhao:2017cud, Zhang:2019jsu, DESI:2024mwx}.
These challenges may suggest anticipated new physics beyond $\Lambda$CDM,
but one thing that is more important before exploring candidate new physics is to
confirm the existence or clarify the source of these challenges.

GW detections allow us to directly estimate the luminosity distances of compact binary \ac{GW} sources to us
without the need for calibrations of the cosmic distance ladder, and the combination of
luminosity distance information from \ac{GW} detections and redshift information obtained by other means
can be used to independently probe the expansion history of the Universe, so \acp{GW} are known as {\it standard sirens} \cite{Schutz:1986gp, Markovic:1993cr, Holz:2005df}.
The LIGO \& Virgo network first realised the idea of probing the cosmic expansion history using standard sirens
through their detected \ac{GW} events \cite{LIGOScientific:2017adf, DES:2019ccw, LIGOScientific:2019zcs}.
The network of current ground-based \ac{GW} detectors has measured $H_0$ with a precision
of about $10\%$, but has yet to provide an effective constraint on the equation of state of dark energy
 \cite{LIGOScientific:2021aug}, thus more \ac{GW} detection plans are needed to shed light on
the challenges faced by $\Lambda$CDM through \ac{GW} standard siren probes.
Space-based \ac{GW} detectors in milli-Hertz band are expected to be launched in the 2030s
 \cite{LISA:2017pwj, TianQin:2015yph, TianQin:2020hid, Hu:2017mde},
the types of candidate sources that
can be detected will be more diverse \cite{LISA:2022yao, Li:2024rnk}, so that the constraints on
the cosmic expansion history from different types of candidate \ac{GW} sources can be complementary and
calibrated with each other \cite{LISACosmologyWorkingGroup:2022jok},
thus space-based \ac{GW} detections will be able to provide irreplaceable roles in
clarifying the Hubble tension and in probing the nature of dark energy.

TianQin can detect several types of candidate standard sirens, namely,
\ac{SBHB} inspirals \cite{DelPozzo:2017kme, Muttoni:2021veo, Zhu:2021bpp},
\acp{EMRI} \cite{MacLeod:2007jd, Laghi:2021pqk, Zhu:2024qpp}
and \ac{MBHB} mergers \cite{Petiteau:2011we, Tamanini:2016zlh, Zhu:2021aat}.
The sensitive frequency band of TianQin is a bit higher than that of LISA, and there is a long baseline between the detectors of TianQin and LISA,
so TianQin and LISA can complement each other well \cite{Shuman:2021ruh, Gao:2024uqc, Liu:2020eko, Lyu:2023ctt, Fan:2020zhy, Torres-Orjuela:2023hfd}.

In this subsection, we will briefly introduce the potential of TianQin for
probing the expansion history of the Universe, including measuring the parameters of the $\Lambda$CDM model and the equation of state of dark energy.
A more thorough discussion of the results can be found in \cite{Luo:2025tqr}.

\subsubsection{Constraining the $\Lambda$CDM model}  \label{sec:sci_cosmo_LCDM} \

The three types of candidate standard sirens for TianQin --- \acp{SBHB}, \acp{EMRI}, and \acp{MBHB}, are
distributed at low ($z<0.3$), medium ($z \lesssim 2$), and high ($z \lesssim 10$) redshifts, respectively,
forming a {\it probe ladder} to better constrain the $\Lambda$CDM model.
For \ac{SBHB} inspirals and \acp{EMRI}, current studies generally assume that they have no observable
electromagnetic (EM) counterparts, requiring the statistical determination of their redshift information by
cross-matching them with surveyed galaxy catalogs
 \cite{DelPozzo:2017kme, Muttoni:2021veo, MacLeod:2007jd, Laghi:2021pqk}.
For \acp{MBHB}, the presence or absence of observable EM counterparts, as well as
their characteristics, depend heavily on the astrophysical environments
surrounding the \ac{GW} source \cite{Petiteau:2011we, Tamanini:2016zlh, Bogdanovic:2021aav}.

Since the low redshift \ac{GW} sources detectable by TianQin include all three types of \ac{SBHB}, \ac{EMRI} and \ac{MBHB} sources, all three types of sources are effective in constraining $H_0$ \cite{Zhu:2021bpp, Zhu:2024qpp, Zhu:2021aat}.
The precision with which TianQin constrains $H_0$ using the \ac{SBHB} detections is expected to be around $20\%$,
potentially improving to about $8\%$ with the TianQin I+II configuration \cite{Zhu:2021bpp}.
These precisions are comparable to those achieved by the current network of ground-based \ac{GW} detectors \cite{LIGOScientific:2021aug},
with the level of precision primarily limited by the low SNR of \ac{SBHB} inspiral signals.
If TianQin and the next generation
ground-based \ac{GW} detectors (such as Einstein Telescope and Cosmic Explorer \cite{Punturo:2010zz,
LIGOScientific:2016wof}) can form a multi-band network, a constraint on $H_0$ of $1\%$ precision level
can be achieved using only about one hundred \acp{SBHB} \cite{Zhu:2021bpp}.
The expected precision with which TianQin can constrain $H_0$ using \ac{EMRI} detections is about $1\%-8\%$ \cite{Zhu:2024qpp}.
Such large uncertainties arise primarily from our limited understanding of the \ac{EMRI} event rates,
which are affected by factors such as the population of massive black holes at the centers of galaxies,
stellar cusps surrounding the massive black holes, the mechanisms by which massive black holes capture stellar-mass objects \cite{Babak:2017tow, Amaro-Seoane:2012lgq},
and activities of the massive black holes \cite{Levin:2006uc, Pan:2021ksp, Pan:2021oob}.
The expected precision for TianQin to use \ac{MBHB} signals to constrain $H_0$ is about $1.5\%-6\%$ in the optimistic scenario and about $2\%-7\%$ in the conservative scenario \cite{Zhu:2021aat,Wang:2019tto}.
The uncertainties in the predictions are again caused by the uncertainty in the \ac{MBHB} event rates \cite{Klein:2015hvg, Wang:2019ryf}.

The prospect of using TianQin to constrain other $\Lambda$CDM model parameters, such as the fractional total matter and dark energy density parameters $\Omega_M$ and $\Omega_{\Lambda}$, is mainly rooted in the \ac{EMRI} and \ac{MBHB} detections \cite{Zhu:2024qpp, Zhu:2021aat}.
Regardless of the type of \ac{GW} sources used, both the constraints of TianQin on $\Omega_M$ and $\Omega_{\Lambda}$
are expected at the dozens of percent level, reaching the 10\% level only in very optimistic scenarios \cite{Zhu:2024qpp, Zhu:2021aat}.
Although the precision on $\Omega_M$ and $\Omega_{\Lambda}$ is not very high here, doing a simultaneous estimation of these parameters can avoid the biases in $H_0$ estimations.

Two of the most representative $H_0$ measurements signifying the Hubble tension are:
$67.4 \pm 0.5 ~\!{\rm km/s/Mpc}$ inferred by the Planck Collaboration with cosmic microwave background (CMB) observations plus the $\Lambda$CDM model \cite{Planck:2018vyg},
and $73.04 \pm 1.04 ~\!{\rm km/s/Mpc}$ obtained by the SH0ES Team with Type Ia supernova (SN Ia) observations plus cosmic distance ladder \cite{Riess:2021jrx},
corresponding to relative precisions of about $1.4\%$ and about $0.7\%$, respectively.
To help break the draw, TianQin will need to provide an independent $H_0$ measurement with a precision better than about $2\%$.
This is possible with the three types of standard sirens, \ac{SBHB}, \ac{EMRI} and \ac{MBHB}, but with some strings attached:
using \ac{SBHB} detections is dependent on the implementation of a multi-band network \cite{Zhu:2021bpp},
and using \ac{EMRI} or \ac{MBHB} detections is dependent on their event rates \cite{Zhu:2024qpp, Zhu:2021aat}.
To help improve the constraints on $H_0$ with a given set of \ac{GW} signals, it is necessary to optimise the data processing process.
Some candidate optimisation methods include:
(i) for \acp{SBHB}, weighting candidate host galaxies with photometric information in all bands
contained in the galaxy catalogs \cite{Zhu:2021bpp};
(ii) for \acp{EMRI}, inferring the correlation between \acp{EMRI} and \acp{AGN} by statistical methods \cite{Zhu:2024qpp};
(iii) for \acp{MBHB}, weighting candidate host galaxies with the correlation between the massive black hole masses and bulge luminosity of galaxies \cite{Zhu:2021aat};
(iv) for lensed \acp{MBHB}, extracting the redshift information of \ac{GW} sources directly using the strong gravitational lensing effect \cite{Huang:2023prq};
and (v) for all three types of \ac{GW} sources, decreasing the spatial localization error volume
by the combined data analysis of TianQin and other detectors \cite{Gao:2024uqc,
Liu:2020eko, Lyu:2023ctt, Fan:2020zhy, Torres-Orjuela:2023hfd, Wang:2019ryf}.
These methods can improve the precision of $H_0$ ranging by a few percent to several times \cite{Zhu:2021bpp, Zhu:2024qpp, Zhu:2021aat, Huang:2023prq}.

\subsubsection{Probing the equation of state of dark energy }  \label{sec:sci_cosmo_DE} \

Dark energy drives the transition of the cosmic expansion from deceleration to acceleration.
A typical phenomenological dark energy model is the \ac{CPL} model \cite{Chevallier:2000qy, Linder:2002et}, which has two parameters, $w_0$ and $w_a$, describing the redshift-independent constant part and the redshift-dependent variable part of the dark energy equation of state, respectively.
The cosmological constant is the simplest dark energy model and the current standard in the $\Lambda$CDM model,
corresponding to a constant equation of state of $w \equiv -1$ (i.e., $w_0=-1$ and $w_a=0$ in the \ac{CPL} model) \cite{Peebles:2002gy, Carroll:2000fy, Li:2011sd}.
The latest observations of the dark energy equation of state published by the Dark Energy Spectroscopic Instrument (DESI) shows a significant deviation from the cosmological constant,
with the most significant deviation coming from a combined constraint of CMB, BAO and SN Ia observations, yielding $w_0 = -0.73 \pm 0.07$ and $w_a = -1.1 \pm 0.3$ \cite{DESI:2024mwx}.
The corresponding relative precision of the parameter $w_0$ is about $10\%$.

Constraining the equation of state of dark energy requires combining observations at different redshifts.
Thus, TianQin's ability to constrain the dark energy equation of state primarily relies on \ac{EMRI} and \ac{MBHB} events whose distributions span from low to high redshifts \cite{Zhu:2024qpp, Zhu:2021aat}.
TianQin's expected precision for constraining the parameter $w_0$ using \ac{EMRI} detections is $|\Delta w_0/w_0| \sim 5\%-40\%$,
but obtaining an effective constraint on $w_a$ is challenging \cite{Zhu:2024qpp}.
Using \ac{MBHB} detections, TianQin can achieve effective constraints on both $w_0$ and $w_a$ in the optimistic scenario with EM counterpart observations,
with expected precisions of $|\Delta w_0/w_0| \sim 7\%-14\%$ and $\Delta w_a \sim 0.4-0.6$ for $w_0$ and $w_a$  \cite{Zhu:2021aat}.
However, in the conservative scenario without EM counterpart observations,
TianQin can only constrain $w_0$, with an expected precision of $|\Delta w_0/w_0| \sim  10\%-40\%$.
Similar to the constraints on $H_0$,
the uncertainties in the expected constraints on the parameters $w_0$ and $w_a$ by TianQin are also
due to the uncertainties in the population models of \acp{EMRI} and \acp{MBHB} \cite{Zhu:2024qpp, Zhu:2021aat}.
As with the constraints on $H_0$, optimising data processing method can also improve the precisions on the equation of state of dark energy \cite{Zhu:2024qpp, Zhu:2021aat}.

\section{Concept and design of TianQin}

{\it Coordinator: Xuefeng Zhang}

\begin{figure}
\centering
\includegraphics[width=0.8\linewidth]{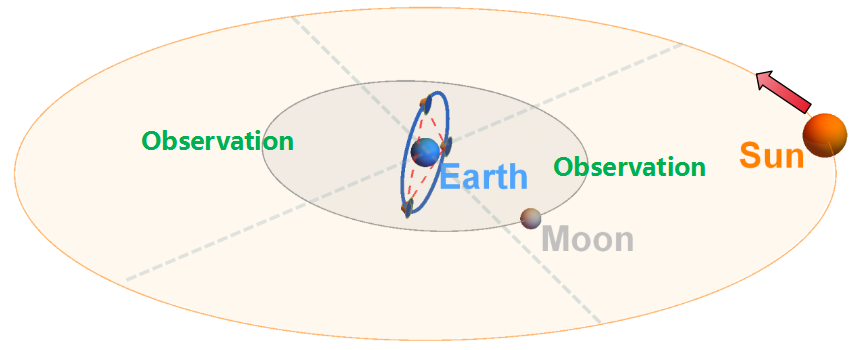}
\caption{The TianQin constellation viewed in the Earth-centered celestial frame (not to scale). The dashed lines mark the two 3-month observation windows when the Sun is roughly facing the orbital plane. }
\label{fig_TQ}
\end{figure}

The concept of TianQin is to deploy three satellites in circular high Earth orbits with a radius of about $10^5$ km, forming a nearly equilateral-triangle constellation. The satellites exchange laser interferometric links to detect \acp{GW} in the mHz frequency band. The constellation plane is nearly fixed and set almost vertical to the ecliptic, facing the verification source RX J0806.3+1527. Due to varying solar angles relative to the constellation plane, the operation scheme consists of having two separate 3-month observation windows within a year when the Sun is roughly facing the constellation plane \cite{TianQin:2015yph} (see Fig. \ref{fig_TQ}).

TianQin features a geocentric concept. This choice of orbits has far-reaching influence on the mission and system design. The design needs to take into account several main aspects and their interplay, these including science objectives, orbits and constellation, space environment, science payload, satellite platform, etc. At the time of the first proposal \cite{TianQin:2015yph}, a well-rounded solutions applicable to TianQin was not present in the literature.

\subsection{Special challenges}

Being close to the Earth has major benefits, e.g., launch and deployment, data communication, telecommand, navigation, etc. These can help mitigate engineering risks and technical difficulties. However, there have also been serious concerns and challenges to the feasibility of geocentric missions in general (see, e.g. \cite{NASA2012}). We name the main five of them:
\begin{itemize}
\item Constellation stability;
\item Eclipses due to the Moon and Earth;
\item The Earth and Moon's gravity disturbance to science measurements;
\item Varying thermal environment; and
\item Solar-wind plasma disturbance to science measurements.
\end{itemize}
In the following, we will briefly review the answers or solutions to these issues, based on the intensive and focused studies carried out so far \cite{Ye:2019txh,Ye:2024dca,Zhou:2021psj,Tan:2020xbm,Zhang:2020paq,Luo:2022pgc,Jiao:2023grf,Liu:2022zue,Zheng:2022put,ZXF2018,Chen:2021dzg,Wang:2024ziz,Lu:2021raf,Su:2021zmj,Jing:2022aaj,YaNan:2024xan},
which have also been recently summarized in \cite{ZXF2024conference}.

\subsection{Orbit and constellation}

Gravitational perturbations from celestial bodies such as the Moon and Sun, along with orbit control errors, cause the \ac{S/C} constellation to deviate from its nominal equilateral triangle configuration.
There are three key performance indices of stability: variations in armlengths, relative velocities, and breathing angles.
The stability of the constellation is crucial and affects several aspects of science payload design such as laser pointing control, phase measurement, laser frequency noise removal, payload architecture, etc. To alleviate pressure on instrumentation, it would be desirable to find orbits with deviation from the equilateral triangle as small as possible. Through optimization techniques for pure-gravity orbits, the constellation can be stabilized to armlength variations within $\pm 0.1\%$, relative velocity changes within $\pm 4$~m/s, and breathing angle fluctuations within $60\pm 0.1^\circ$ over two years \cite{Ye:2019txh,Ye:2024dca}. The natural fluctuation of breathing angles caused by gravitational perturbations is estimated to be $\pm 0.06^\circ$, leaving margins for orbit determination and control. Additionally, the detector pointing is quite stable, varying by $<3^\circ$ over five years. Moreover, the implications for \ac{TDI} have been studied, and it indicates that the first-generation \ac{TDI} is applicable to TianQin due to a small residual armlength mismatch \cite{Zhou:2021psj}, which can make data processing on the ground easier.

Eclipse avoidance is important for TianQin due to potential thermal shocks and power disruption to the satellites by traversing Moon and Earth's shadows. This was considered a major risk for geocentric concepts \cite{NASA2012} with remedies not well studied. To address the issue, an eclipse-free orbit design was proposed to ensure that the satellites do not undergo Moon eclipses during the observation windows of 3 + 3 months \cite{Ye:2020tze}. Although Earth eclipses are unavoidable, they occur outside these windows, hence less worrisome. The design proposes 1:8 synodic resonant orbits with respect to the Moon, which is to have the satellites to complete eight revolutions in one lunar month. The idea is to set the satellite motion in a repeated phase relation with the Moon's shadow to reduce the possibility of Moon eclipses. An optimal orbital radius of 100,935 km has been identified for a launch in the early 2030s. The resulting optimized orbits can avoid all eclipses in 3+3 months observation windows for a 5-year mission and meet the constellation stability requirements. Additionally, eclipse-free observation windows can be further extended to 4+4 months at an acceptable expense of reduced allowed ranges of initial phases of the satellites.

The mission orbit design needs to take into account and balance among a variety of factors such as science objectives, constellation stability, space environment, science payload, launch capability, etc. Each design factor may have its own preference. Based on our current trade-offs, the mission orbit is proposed to have a semimajor axis of $a = 100935$ km, eccentricity $e < 0.002$, inclination $i = 94.7^\circ$, and ascending node $\Omega = 210.4^\circ$ in Earth-centered ecliptic coordinates \cite{Tan:2020xbm,Ye:2020tze,Luo:2022pgc,Zhang:2021sys}. To keep the constellation stable for at least 3 months, typical initial orbit error requirements are $\pm20$ m in the radial position and $\pm2$ mm/s in the along-track velocity, which will be addressed in Sec. \ref{sec:sat}. The three satellites are foreseen to be delivered by one single heavy-lift launch vehicle into an elliptical transfer orbit, then reaching their final orbits by raising the perigees.

\subsection{Space environment}

Gravitational disturbances from Earth and Moon may significantly affect the geodesic motion of \ac{TM} in the inertial reference systems, potentially inducing ``orbital noise'' similar to Newtonian noise observed in ground-based detectors. To assess the ``quietness'' of the ambient gravitational field, we have chosen the range accelerations between \acp{TM} to perform the evaluation. The criteria for this assessment involve comparing the amplitude spectral density of the residual acceleration noise between two \acp{TM}, with the requirement of $\sqrt{2} \times 10^{-15} \, \mathrm{m/s}^2/\sqrt{\mathrm{Hz}} $ in the frequency range of $ 10^{-4} \, \mathrm{Hz} $ to $ 1 \, \mathrm{Hz} $. Achieving this precision necessitates the use of quadruple precision to propagate pure-gravity orbits. From careful modeling and simulation, the total gravitational disturbance is expected to dominate only at lower frequencies and drop below the requirement at $ 1 \times 10^{-4} \, \mathrm{Hz} $, hence not presenting a showstopper to the mission \cite{Zhang:2020paq,Jiao:2023grf,Liu:2022zue}. By positioning the satellites at an orbital radius greater than $1\times 10^5 \, \mathrm{km} $, these disturbances can be pushed out of the sensitive frequency band \cite{Luo:2022pgc}.

The thermal environment is crucial for the performance of inertial sensors and laser interferometers, necessitating stringent requirements of the level $10^{-5}\, \mathrm{K}/\sqrt{\mathrm{Hz}}$. A primary challenge for TianQin stems from the varying direction of sunlight relative to the constellation plane, which can introduce thermal variations that may impact instrument performance. To mitigate these effects, a flat-top sunshield design was proposed \cite{ZXF2018}, adapted from science missions such as WMAP, Gaia, JWST, LISA, and LPF. The design ensures that no other parts of the satellite are directly exposed to sunlight, allowing steady heat pathways from the top to the bottom and sides of the satellite. Consequently, the design effectively shifts solar flux variations to a half-year period, minimizing their influence on the instruments. Within the detection band, fluctuations of the solar constant dominate for TianQin \cite{Chen:2021dzg}. This underscores the importance of robust thermal control for maintaining the desired operational temperature range and stability, which is critical to the overall success of the mission.

The interaction of solar-wind plasma with laser propagation between the satellites was suspected to induce extra phase noise, affecting the precision of interferometric measurements. To evaluate this effect, magnetohydrodynamic simulations have been conducted using the Space Weather Modeling Framework along with observational data from the OMNI database \cite{toth2005space}. The results indicate that the induced phase noise stays below the allocated noise budget of approximately 30\% of the total noise, except during strong solar storms that can raise the level \cite{Lu:2021raf,YaNan:2024xan}. Furthermore, it was found that the application of \ac{TDI} combinations can effectively mitigate the noise effect. Specifically, \ac{TDI} is able to achieve about 50\% noise suppression in the frequency range below $10^{-2}$ Hz. This noise suppression is primarily due to the noise correlation among the three arms of the interferometric setup \cite{Su:2021zmj,Jing:2022aaj}. Thus, the influence of solar-wind plasma is fairly understood for the operation and performance of TianQin.

\subsection{Science payload}

The choice of the payload architecture and pointing control strategy is crucial, particularly considering the Earth-Moon environment faced by TianQin \cite{Fang:2024odm}.
The core payload comprises the telescope, optical bench, and inertial sensors, rigidly connected to one another, and the in-plane pointing is achieved via single-axis rotation of the optomechanical assembly, while off-plane pointing is managed through satellite attitude control \cite{LISA:2017pwj}.
The alignment of the optomechanical assembly axes and \acp{TM} with respect to received beams is facilitated by differential wavefront sensing to achieve an accuracy in the order of $10 \, \mathrm{nrad}$. The coupling of orbit and attitude needs careful consideration, as changes in orbit affect the pointing and attitude control. Under breathing angles of $60 \pm 0.1^\circ$ and constellation plane variations of $\sim 0.05^\circ$ per orbit, the nominal force and torque for \ac{TM} suspension control are estimated to be well below the budget, where the acceleration noise requirement mandates the control acceleration below $10^{-10} \, \mathrm{m/s}^2$ and angular acceleration below $10^{-10} \, \mathrm{rad/s}^2$. Moreover, control forces can be minimized by positioning of the satellite’s center of mass at the midpoint between \acp{TM} \cite{Fang:2024odm}.

The point-ahead angle control is a critical aspect in accurate laser pointing between satellites by addressing delays from the finite speed of light. The point-ahead angle, defined as the angle between the received and transmitted beams, is fully determined by the satellite orbits. Orbit simulation for TianQin has established an in-plane static bias of 23 \textmu{}rad, with variations of $\pm 25$ nrad in-plane and $\pm 10$ nrad off-plane. Hence, a fixed-value compensation strategy can be adopted to absorb small and slow variations into pointing biases up to $\pm 35$ nrad \cite{Wang:2024rxl}. From the far-field \ac{TTL} coupling with a pointing jitter of 10 nrad/$\sqrt{\mathrm{Hz}}$, the requirement on far-field wavefront quality has been derived accordingly. A preliminary \ac{TTL} calibration procedure using null \ac{TDI} channels has been put forward and assessed via numerical simulation \cite{Wang:2024tis}. The strategy has major benefits in simplifying interferometer design, payload operation, and \ac{TTL} noise mitigation.

To help with performance assessment and noise budgeting, the team has also developed \ac{TDI} and data pre-processing simulation tools (TQTDI) \cite{Zheng:2022put}.
The simulation utilizes sub-pm/$\sqrt{\mathrm{Hz}}$ orbits from TQPOP \cite{Zhang:2020paq} and multi-body attitudes from TQDYN \cite{Fang:2024odm,ZHANG202457} to generate more realistic heterodyne beatnote signals, capturing the complex dynamics of the satellites and the constellation. Additionally, to remove Doppler shift due to Earth-Moon's gravitational disturbances, a high-performance high-pass filter is implemented, which is shown to be compatible with \ac{TDI}. These techniques enhance simulation fidelity and aid in more accurate analyses.

\subsection{Satellite platform} \label{sec:sat}

{\it Coordinator: Xuefeng Zhang, Ming Li, Lihua Zhang}

The satellite platform is engineered to fulfill strict requirements in thermal, magnetic, self-gravity, structural, and vibrational aspects, which requires a customized design that addresses both external and internal environmental challenges in space (see, e.g. \cite{ZSDZ2021Z1013}). The preliminary configuration features a single flat-top sunshield, thermally isolated to provide over three months of effective shading \cite{ZXF2018}. The octagonal squashed outer shape, along with central support and shear panels, helps to enhance structural rigidity \cite{zhang2023p,Wang:2024ziz}. Additionally, the compartmentalized and symmetric layout aids in mass balancing and self-gravity mitigation.

The thermal design of the satellite aims to maintain stable operating temperatures despite significant solar angle and thermal flux variations that can reach approximately 38\% over three months. Following passive thermal guidelines, the preliminary design incorporates a sunshield and top-plate equipped with optical solar reflectors, polyimide foams, gold coatings, and aerogel for enhanced insulation. The simulation results show that the key payload bay can maintain a slight variation of $<2$ K over three months \cite{Wang:2024ziz}. Due to the importance of thermal control, further discussion is deferred to a separate subsection (Sec. \ref{sec:thermal}).

The satellite employs cold gas thrusters to execute drag-free operations and precise pointing control, particularly under varying solar radiation pressure, both seasonally and in orbital periods. To avoid plume impingement, the thruster directions are limited, and they are oriented downward and away from the sunshield and satellite body, with solar radiation pressure as a virtual thruster. Both 3-cluster and 4-cluster designs have been tested \cite{Fang:2024odm}. Allocation of the required total force and torque indicates positive-output solutions can be achieved for four months in the science mode, ensuring consistent drag-free and pointing accuracy throughout the mission. Furthermore, electric propulsion with higher thrusts is also foreseen to assist precise orbit control during non-observation periods.

Other features of the satellites include \ac{GNSS} receivers and real-time data downlink. The former is adopted for precise orbit determination using \ac{GNSS} leak signals, which has the advantages of reducing ground-based tracking infrastructure costs and receiving real-time data \cite{ZHANG20253050}. Previously tested in Chang’e missions, it is estimated that the method has an accuracy of 4.8 m in radial position and 0.2 mm/s in along-track velocity over a seven-day arc \cite{TONG2025534}, which can meet the requirement. \ac{GNSS} can also be used jointly with satellite laser ranging  \cite{An:2022xsi,Zhang:2022gsz} and other methods for further improvement \cite{An:2024hnu}.
For data communication, a high-gain antenna is to be installed to the front panel of the satellite and can remain fixed and automatically Earth-pointing in the science mode, and provide a data rate greater than 1 Mb/s. Additionally, data can be relayed via laser links to at least one satellite visible to the ground, enabling continuous real-time downlink to support quick alerts of important merger events, thereby enhancing mission responsiveness \cite{Yi:10444526}.

\subsection{Thermal control}\label{sec:thermal}

{\it Coordinator: Ran Wei, Xin Zhao}

The thermal control subsystem is an important component of the entire satellite. To achieve the \textmu{}K-level temperature control required by the core payload, developing the thermal control subsystem needs research on mechanism for temperature noise transmission and attenuation, high-resolution temperature measurement theory, and ultra-low noise active suppression methods, breaking through core technologies such as temperature noise suppression, high-precision temperature measurement sensing, and test verification.

Research has been carried out on the temperature noise suppression technology exploring the transmission and attenuation mechanisms of different frequency disturbances in the thermal structure.
It constructs a multi-level temperature control system model using control equations based on time constants and validates it with a damping oscillation model based on the theory of heat and mass, providing a design method for multi-level energy attenuation control.
Combining the configuration layout, flight attitude, and orbital characteristics of TianQin, a high-precision thermal network model is constructed using a hierarchical approach, and some preliminary result on the energy flow and temperature distribution of the satellite has been obtained. Through comparative analysis of the finite difference thermal network model and the multi-coupled thermal effect model of the high vacuum ($\leq10^{-6}$ Pa) region around the inertial sensor \ac{TM},
the frequency-domain thermal analysis results around the inertial sensor \ac{TM} are obtained \cite{HAO2024S,ZHANG2024ATE,WANG2024JAP,SUN2024JPCM,ZUO2023TI,SUN2023IJHMT,SUN2023NT}.
The results show that the temperature noise of the \ac{TM} part can be controlled below 5 \textmu{}K/Hz$^{1/2}$.

In engineering, to achieve \textmu{}K-level temperature measurement, the ultra-low-frequency noise estimation and separation technology based on multi-channel cross-spectrum circuits and the isolation power supply and bidirectional drive differential proportional ultra-high-resolution temperature measurement technology have been developed. In the frequency band of 0.1 mHz to 1 Hz, the equivalent temperature noise spectral density in the laboratory has reached 9 \textmu{}K/Hz$^{1/2}$, and the goal of 2 \textmu{}K/Hz$^{1/2}$ is being pursued \cite{2023FrP....1112368B,2023AAXAA,2024Application,2023JPhCS2617a2001W,2024NatSR..1410802B,2025Meas..24015538P,2025OptLT.18111999M,Tong2024ass}. To eliminate the influence of long-period thermal flow disturbances, a low-frequency temperature noise suppression technology based on feedback correction phase-shift compensation algorithm has been developed, significantly reducing the impact of long-period disturbances \cite{YANG2024123767,MAO2024125512,DENG2024108021,HAO2024152427,Chen:2024iit}. Combining the above two technologies, a high-precision temperature controller has been developed, and the principle prototype design have been completed.

According to the overall satellite configuration and thermal design plan, a high thermal resistance and high thermal capacity structure is used under the solar cell array to attenuate and isolate the space thermal flow. With the technology to create high-gel-activity sol, two types of thermal insulation materials with high strength and low modulus and ultra-low thermal conductivity have been prepared. The thermal insulation material with thermal conductivity $<0.03$ W/(m·K) has been developed \cite{Zhang2024japs,zhang2024mat}.

For design verification, the test schemes for the scale model and the fast response model have been determined, and the simulation method corresponding to the satellite's external heat flow has been formulated, through the  equivalent simulation technology with frequency-domain discretization of temperature boundary and heat flow boundary \cite{10.1007/978-981-97-2120-7_8}, and by formulating the sinusoidal amplitude response curve based on the transfer function. The preliminary design of the corresponding tests has been completed.

\section{Developing Key Technologies for TianQin}

{\it Coordinator: Zebing Zhou, Hsien-Chi Yeh, Chao Xue}

In this section we report the progress on the two key technologies of the TianQin project: the inertial reference technology and the inter-satellite laser interferometry technology.

\subsection{Inertial reference}

{\it Coordinator: Zebing Zhou}

In the context of space-based \ac{GW} detection, the inertial reference is of paramount importance. The system is comprised of three fundamental components:
the inertial sensor is meticulously engineered to minimize \ac{TM} disturbances through components such as the sensitive head and sensing/control systems;
the micro-Newton thrusters are developed to counteract non-conservative forces acting on the satellite platform;
drag-free control is achieved through dynamic modeling, diverse control algorithms, and ground simulation experiments, ensuring precise satellite and \ac{TM} motion control.
These elements collectively contribute to the success of the detection mission.

\subsubsection{Inertial sensor}\

{\it Coordinator: Yanchong Liu}

Based on functions and disturbance suppression requirements, inertial sensors are mainly composed of the sensitive head, sensing and control system, charge management system, caging and releasing system, vacuum maintenance system.
For the space-based \ac{GW} detection, the \ac{TM} is the inertial reference and provides a reference point for laser interferometry.
TianQin requires the residual acceleration of the \acp{TM} along the sensitive axis to be less than $1\times10^{-15}\ \mathrm{m/s^2/Hz^{1/2}}$ \cite{TianQin:2015yph}.

\paragraph{Sensitive head}

The sensitive head is the core of the inertial sensor, which is composed of \ac{TM}, electrode plate and electrode housing.
The \ac{TM} must have high density, superior thermal conductivity, low remanence, and low magnetic susceptibility to minimize disturbances.
In TianQin, the \ac{TM} is a cubic gold-platinum alloy with a mass of about $2.5$ kg.
Considering the low disturbance requirement of the sensitive axis and carrier injection requirement, the plate configuration scheme of three-axis control plate and two-non-sensitive axis injection plate will be used for TianQin.
The electrode housing is responsible for carrying the plate and providing the necessary electromechanical interface for the other system.

Prototypes of the \ac{TM} and electrode housing have been created.
The \ac{TM} exhibited excellent magnetic performance, with a susceptibility of $8\times 10^{-6}$ and a residual magnetic moment of approximately $10~\mathrm{nA\cdot m^2}$, meeting the requirement of TianQin \cite{louTheoretical2023}.

\paragraph{Sensing and control system}

The sensing and control system mainly includes capacitance sensing unit, electrostatic control unit and control unit.
The capacitive sensing unit and electrostatic control unit are used for the measurement and control of the six degrees of freedom motion of the \ac{TM}.
For TianQin, the requirement for capacitive sensing is $7\times 10^{-7}\ \mathrm{pF/Hz^{1/2}}$ at $6$ mHz.
A prototype of the capacitive sensing unit has been created with a resolution of $2.3\times 10^{-7}\ \mathrm{pF/Hz^{1/2}}$, mainly limited by thermal noise from the transformer and pre-amplifier \cite{yangmethod2023,liCapacitive2024,huResonant2014,Bai:2015tva,baiCapacitive2009}.
A prototype of the electrostatic control unit has been realized with a voltage noise of $1\times 10^{-5}\ \mathrm{V/Hz^{1/2}}$, meeting the requirement of TianQin.

\paragraph{Charge management system}

The charge management system is responsible for measuring the \ac{TM}'s residual charge and using ultraviolet discharge technology to neutralize the residual charge resulting from high-energy charged particle collisions on the \ac{TM}.
A $254$ nm Micro-LED with power exceeding $3 \times 10^{-6}$ W has been developed, with a measured lifetime exceeding $5000$ h.
An engineering model for the charge management subsystem has been developed, with a charging rate capability of $5\times 10^5 ~\mathrm{e/s}$ \cite{Yang:2020dgd,yangcharge2020,liCoupling2021,liCapacitive2024,chenUsing2024}.
A ground test and verification of the charge management system using a torsion pendulum  was constructed, achieving a charge control level of $6.6\times 10^{-14}$ C, which is almost an order of magnitude better than the requirement ($3.4\times 10^{-13}$ C).

\paragraph{TM Lock and release system}

Due to the substantial mass of the \ac{TM} and its considerable separation from the electrode frame, the unconstrained \ac{TM} will possess destructive kinetic energy during the launch phase.
Therefore, to ensure the safe transition of the inertial sensors into the scientific operational mode, a dedicated lock and release system is necessary, taking into account the launch conditions and release requirements \cite{Xue:2024cbs}.
The system is designed to fully lock the \ac{TM} during the launch phase, release it with near-zero initial velocity to a free-falling state, and subsequently capture it using an electrostatic actuator.
The design of the lock and release system has been finalized, and a prototype of the principle has been developed. Preliminary test of the system's locking, capturing, positioning and releasing functions has been completed \cite{SHI2025}. The lock force of the system has been determined to exceed 1500 N, and the transferred momentum and angular momentum have been estimated at the level $10^{-5}$ kg m/s and $10^{-7}$ kg m$^2$/s, respectively.

\paragraph{System integration and ground testing}

A preliminary design of inertial sensor system integration for TianQin has been completed, with titanium vacuum chamber developed, probe and vacuum chamber installed together, and caging and release subsystem tested.
The integrated inertial sensor flight model is planned to be tested with the TianQin-2 satellite.

Regarding disturbance investigation, high precision torsion pendulum has been used to investigate patch effect \cite{Yin:2014oca}, magnetic effect \cite{yinMeasurements2021,songhigh2023}, thermal gradient effect \cite{Li:2024cev}, and gas damping effect \cite{zhaoExperimental2023} thus far.
The experimental results are in agreement with the theoretical analysis.
Design and requirements of the inertial sensor are iteratively revised based on theoretical analysis and experimental results.

\subsubsection{Micro-Newton thruster}\

{\it Coordinator: Peiyi Song, Jianping Liu, Yong Li}

As the actuator of the \ac{DFC} system, the micro-Newton thruster is a key component and the performance of the micro-Newton thruster constrains the level of \ac{DFC} largely. TianQin mandates high requirements for micro-Newton thrusters, including continuously adjustable thrust between 0-100 \textmu{}N, resolution of $<$0.1 \textmu{}N, and thrust noise of $<$0.1 \textmu{}N/Hz$^{1/2}$ \cite{wu2010attitude}.

Cold gas micro-Newton thrusters offer ultra fine precision and superior reliability. The cold gas micro-Newton thrusters used in \ac{GW} detection require three key technologies: micro-Newton piezoelectric thruster, microgram gas flowsensor and low noise thrust control.
The controller converts commanded thrust to flow rate according to on-ground calibrated formula, then the piezoelectric thruster adjusts the throat area of the proportional piezoelectric valve based on the feedback value of the gas flowsensor. This converges the gas flow rate to the setting point, thereby achieving closed-loop control of thrust. In order to match gas flow rate to thrust, the gas flowsensor must be as close as possible to the piezoelectric valve and integrated inside cold gas micro-Newton thrusters. \cite{Ricci:2020zom,noci2009cold,LISAPathfinder:2019eny}.

The performance of the cold gas micro-Newton thruster relies on high-precision piezoelectric-driven flow control method. Piezoelectric-driven elements in the cold gas micro-Newton thruster utilize the inverse piezoelectric effect to convert electrical energy into mechanical energy, which is then transmitted through a specific structure to achieve fluid control and regulation. Due to the characteristics of high precision, high resolution, low power consumption, low electromagnetic interference, and relatively mature technology, piezoelectric-driven valves have a broad application prospect in the field of high-precision fluid control and have become the preferred solution for the cold gas micro-Newton thruster of space-based \ac{GW} detection missions \cite{mel2018kinematics,zhang2024high}.

Due to the extremely low thrust of cold gas micro-Newton thrusters, they are highly susceptible to the flow characteristics of the propellant and the flow related disturbances, making it difficult to achieve high-precision and stable thrust output through open-loop control. It is necessary to establish an accurate thrust model and closed-loop thrust control methods. The flow sensor acts as the reference signal for thrust control, with its flow measurement resolution setting the upper threshold for thrust control performance. Employing a thermal flow sensing solution, the flowsensor leverages the advantage of temperature differential measurement to mitigate environmental interferences. \cite{DayaBay:2015kir}.

Targeting the need of the TianQin project, research on micro-Newton cold gas thrust control technology has been carried out and in-orbit verification has been achieved with the TianQin-1 satellite.
The micro-Newton gas thruster carried by TianQin-1 has a thrust resolution of approximately 0.1 \textmu{}N. The in-orbit experimental measurements show that the thrust can be precisely adjusted from 1 to 60 \textmu{}N.
For periods where the output of the micro-Newton thruster remains constant, the overall thrust noise is 0.3 \textmu{}N/Hz$^{1/2}$  at 0.1 Hz \cite{Luo:2020bls}.

The construction of the micro-Newton gas propulsion system for the TianQin-2 is being carried out as planned.
It consists of a gas bottle, a gas addition and exhaust valve, a self-locking valve, a high-pressure pressure sensor, a low-pressure pressure sensor, a filter, a micro-Newton variable thrust module, a circuit box, and piping components.
The expected thrust range is 0.1 to 1000 \textmu{}N.

\subsubsection{Drag-free control}\

{\it Coordinator: Guoying Zhao, Xingyu Gou}

In this subsection, the current status of \ac{DFC}  technology in the TianQin project is briefly summarised from the following three perspectives: dynamic modelling, control algorithms, and ground simulation experiments.

\paragraph{Dynamic modelling}

The main objective of the dynamic modelling is to obtain mathematical models that describe the motion coupling between the satellite, the optomechanical assembly and the \acp{TM}. In addition, it is a fundamental step to derive linearised and decoupled models and to develop suitable controllers.

The \ac{DFACS} can be divided into three main phases after the \ac{S/C} reaching orbit: releasing \ac{TM}, constellation acquisition, and scientific mode. The interaction between the \ac{S/C} attitude, laser pointing, and \ac{TM} leads to complex multi-degree-of-freedom control across multiple modes. In addition, the core payload is currently in the phase of being systematised. It is therefore necessary to investigate the impact of different system configurations on the control system. These requirements demand a sufficient understanding of the \ac{S/C} dynamics model.

At present stage, a dynamic model has been developed that can completely describes the kinematics of the motion coupling. Compared to other models, this model does not have restrictive assumptions (e.g. kinematic states and system parameter configurations). The accuracy of the dynamic model has been verified by comparing it with the Simscape model. The results show that the model has higher accuracy than the existing model in all degrees of freedom.
Monte Carlo simulations have been performed using random parameters to demonstrate that the model can be used for different parameter configurations and three different stages of the \ac{DFACS}. What's more, a corresponding simplified model in scientific mode is proposed based on the orbital configuration of the TianQin \ac{S/C}. The simplified model greatly reduces the system complexity while maintaining the main system dynamics, and therefore can further improve the efficiency of control design in scientific mode. In summary, the obstacles of the \ac{DFACS} in modeling the dynamics of \ac{S/C} have been removed. More detailed information about the dynamic model can be found in \cite{ZHANG202457}.

\paragraph{Control algorithm}

Control algorithms are the core of \ac{DFACS}. A variety of control algorithms have been developed now.

A simple single integral controller has been implemented for the acceleration-mode \ac{DFC} on TianQin-1, focusing on solving practical engineering problems and achieving stable, reliable, and highly repeatable control results \cite{gou2021acceleration}.
For \ac{DFC} in general, the displacement-mode DFC feasibility and thrust saturation avoidance has been explored \cite{gou2024study}, adaptive control theories, proposing adaptive methods, and designing estimation and control strategies have been developed \cite{zou2020characteristic,Wang2021AdaptiveDC,Tan2021Adaptive,yang2020extended}.
A new design strategy has been proposed for the Drag-Free \ac{S/C}, which effectively solves the frequency band limitation of the control signal in the scientific mode through frequency separation theory, while reducing fuel consumption \cite{lian2021frequency}.
A multi-loop controller design method has been  proposed for the Drag-Free \ac{S/C}, which effectively improves the design efficiency by transforming the performance index into frequency domain constraints and using a multi-group genetic algorithm to optimize the controller design \cite{ma2022controller}.
A model predictive control method has been proposed to solve the fault-tolerant control problem of a Drag-Free \ac{S/C} under actuator faults and input saturation \cite{10.1063/5.0136994}.
An embedded model control (EMC) structure has been designed based on a decoupled dynamic model and used a loop shaping method to adjust the controller parameters \cite{xiao2022drag}.
The proposed controller design was verified by numerical simulation.
The results show that the fluctuation of the relative displacement of the \ac{TM} is controlled below $3\ \mathrm{nm/Hz^{1/2}}$, which meets the requirements of TianQin.
In addition, Monte Carlo experiments show that the dynamic performance of the closed loop is only slightly affected by parameter uncertainties, demonstrating the robustness of the EMC.
A proportional-integral-derivative feedforward controller has been designed for TianQin-1 and implemented it in the FPGA chip of the electrostatic accelerometer \cite{10.1063/5.0189320}.
The performance of the \ac{DFC} system was verified through in-orbit experiments.
The results show that a residual acceleration of $5\times10^{-11}\ \mathrm{m/s^2/Hz^{1/2}}$ at 10 mHz was achieved.

Current research results on control algorithms focus mainly on scientific models and neglect the degrees of freedom of the optomechanical assemblies. Control design that considers the complete degrees of freedom and other modes in addition to scientific modes is currently ongoing. For more detailed information on control design, please refer to the references \cite{ZHANG202457, gou2021acceleration, gou2024study, zou2020characteristic, Wang2021AdaptiveDC, Tan2021Adaptive, yang2020extended, Wang2021AdaptiveDC, lian2021frequency, ma2022controller, 10.1063/5.0136994, xiao2022drag, 10.1063/5.0189320}.

\paragraph{Ground simulation experiments}

Given the high cost and significant risk of direct in-orbit testing of the \ac{DFC} system, ground-based simulation and validation are essential.

A high-precision satellite simulator designed to validate inter-satellite laser tracking performance has been developed.
An air-bearing system is employed to minimize friction and torque between the simulator and the underlying surface.
The satellite platform is equipped with four sets of orthogonally mounted gas thrusters, which can be combined to generate the desired forces and torques for the satellite platform.
The satellite platform performs coarse laser tracking, while an optical platform is used for fine laser tracking.

The developed EMC algorithm has been applied to this satellite simulator. Results indicate that the satellite platform achieves measurement tracking errors of better than $10\ \mathrm{mrad/Hz^{1/2}}$ during fixed-position tracking and better than $50\ \mathrm{mrad/Hz^{1/2}}$ during motion tracking. The optical platform reduces measurement tracking errors to 80 \textmu{}rad/Hz$^{1/2}$. This simulator demonstrates potential for simulating laser tracking missions. More detailed information about the satellite simulator can be found in \cite{10.1063/5.0189320}.

TianQin is also currently constructing some other ground simulation facilities that can be used for Drag-Free verification.

\subsection{Intersatellite laser interferometry}

{\it Coordinator: Hsien-Chi Yeh}

The inter-satellite laser interferometry system is the core payload system for measuring the optical-path-length changes caused by the \acp{GW}.
So far, all proposed space-based \ac{GW} detection missions are based on three-satellite constellation with six inter-satellite laser links. Due to a very weak light power received by the remote satellite after propagating through a large inter-satellite distance, a transponder-type interferometer, instead of a typical Michelson’s interferometer, must be used for inter-satellite laser interferometry.

A simplified configuration of inter-satellite transponder laser interferometer consists of a frequency-stabilized laser, ultra-stable optical benches, weak-light phase locking control unit, and telescopes that can be treated as laser couplers to the ultra-stable optical benches.
In TianQin-1 mission, we have tested the performances of ultra-stable optical bench and phasemeter. The noise level of laser interferometer achieved 30 pm/Hz$^{1/2}$ @ 0.1 Hz.
In TianQin-2 mission, planned to be launched around 2026, a complete inter-satellite transponder laser interferometer will be demonstrated. Besides ultra-stable optical benches and phasemeters, a frequency-stabilized laser unit and a weak-light phase locking control unit will be demonstrated in orbit.

In the following, we present the current progress on developing the frequency-stabilized laser, the ultra-stable optical benches, the weak-light phase locking control unit and the telescope.

\subsubsection{Frequency stabilized Laser}\

The laser source is based on a master oscillator power amplifier architecture, and the output frequency is stabilized by the \ac{FSU}.
There are three main parts: the seed laser, the optical amplifier and \ac{FSU}.

The seed laser is a Nd:YAG nonplanar ring cavity solid-state laser pumped by 808-nm laser diodes \cite{WOS:000975312500007,WOS:001018917000006}, which has two tunning channels, the piezoelectric tunning channel and the temperature tunning channel, to realize the functions of frequency stabilization and phase locking.
The free-space light generated by the pump diode is focused into the Nd:YAG crystal through coupling lens, and the output 1064-nm free-space laser is linearly polarized by wave-plates and then injected into a single-mode polarization maintaining fiber through a fiber coupler. In order to improve the reliability of the seed laser, two pump diodes are packaged into the seed laser, while under nominal condition, only one pump diode is working and the other is the offline backup.

The optical amplifier is based on a Yb-doped fiber amplifier architecture, which can output watt-level linear-polarized laser at 1064-nm wavelength \cite{Li:25}. The seed laser and a 976-nm pump laser are injected into the double-cladding active fiber through a fiber coupler. The fiber mode field adapters and fiber isolators are used to adjust the mode size in the fiber and to protect the seed laser. In the end of the fiber optical link, fiber splitters are used to realize power stabilization feedback and power monitoring, while the port on the main path outputs watt-level laser for inter-satellite laser interferometry. Multiple pump diodes are packaged in the optical fiber amplifier to improve the reliability and to compensate for potential laser power degradation during the operation period in space.

The \ac{FSU} adopts the integrated quasi-monolithic ultra-stable optical resonator made by using hydroxide-catalysis bonding technique \cite{WOS:000636685400034,10.1063/5.0224636}, and the laser frequency locking is realized with the Pound-Drever-Hall scheme. The \ac{FSU} use all-fiber optical link and FPGA-based digital feedback controller that performs the frequency locking to the resonance frequency of the ultra-stable cavity.
The measured frequency noise spectral density is about 30 Hz/Hz$^{1/2}$ at Fourier frequencies from 10 mHz to 1 Hz.
When testing the frequency noise, the prototype is actively suppressed from vibration noise on the ground, and uses active temperature control to ensure that its temperature is near the zero-expansion-point of the ultra-stable cavity. The prototype is compared with another set of independent ultra-stable cavity \ac{FSU} (same active vibration isolation with the active temperature control at zero-expansion-point), where the beat frequency is measured by a frequency counter.

The three main parts of laser source (seed laser, optical amplifier and \ac{FSU}) are independently designed and packaged.
The prototypes of the three units have passed typical space-compatibility tests, including random and periodic vibrations, impact, thermal cycles, etc.
In the next step, it is necessary to carry out the integrated design among the three units under full consideration of the space environment adaptability.
In addition, in order to realize clock comparison and pseudo-random-code ranging, an optical modulation module will be added to the integrated design.
The laser power stabilization unit is also under development.

\subsubsection{Optical bench}\

The optical bench is a core payload of the laser interferometer system used in space-based \ac{GW} detection missions.
The design, engineering fabrication, noise analysis and suppression of optical benches are the focus of the state-of-the-art researches \cite{Heinzel:2003uh,Heinzel:2005tf,WOS:000314746600007,Armano:2021cwl,Armano:2022ose}.
For TianQin, the ultra-stable optical bench and FPGA-based phasemeter have been demonstrated in TianQin-1 mission \cite{Luo:2020bls,Ming:2024nok}, reaching the noise level 10 nm/Hz$^{1/2}$ @ 6 mHz and up to 100 nm/Hz$^{1/2}$ @ 1 mHz.
For the measurement noise of interferometer at frequency range of 0.1 Hz $\sim$ 1 mHz, it has been investigated and found that the stray lights reflected from the optical components of the optical bench is one of the major error sources.

The stray light occurred on the optical bench is equivalent to an additional reference laser beam that also generates an interference signal when beating with the received weak light, and then provides an additional phase into the total phase difference measured. The theoretical model reveals that this stray light effect of the optical bench results in a nonlinear error, periodically varying with the \ac{OPL}. Base on the periodic property of this nonlinear optical-path-length error, a proper optical path length can be chosen as the initial measuring position of laser interferometer so that the nonlinear error is minimized, hence the stray light effect can be negligible \cite{WOS:001254870400005}.

\begin{figure}[t]
\centering
\includegraphics[width=\textwidth]{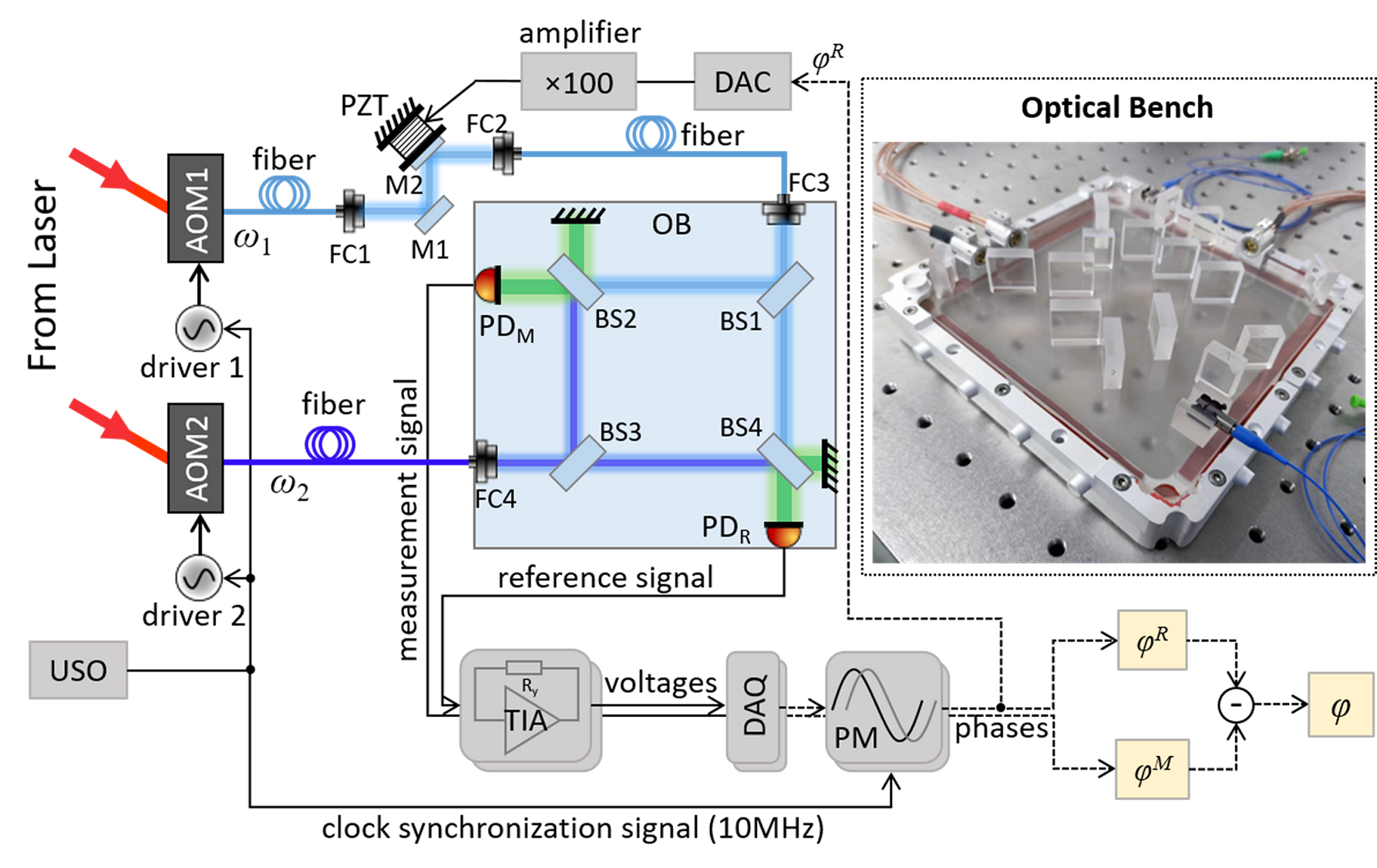}
\caption{Experimental setup of an \ac{OPL}-tunable heterodyne Mach-Zehnder interferometer \cite{WOS:001254870400005}. USO: ultra-stable oscillator; FC: fiber collimator; BS: beam splitter; TIA: trans-impedance amplifier; DAQ: data acquisition device; PM: phasemeter.}
\label{fig:MZI}
\end{figure}

A Mach-Zehnder interferometer system has been built to test the theoretical model and the feasibility of the error suppression method.
Fig. \ref{fig:MZI} shows the experimental setup.
A heterodyne frequency between two laser beams is generated by two acousto-optic modulators (AOM1 and AOM2).
The modulated laser beams are guided into a Mach-Zehnder interferometer that is made with silicate bonding technique.
In order to adjust the \ac{OPL} of the reference beam of interferometer, a PZT-driven mirror, M2, is installed in the optical path of the upper laser beam.
Two interference signals are detected by photodetectors, PDM and PDR, and then are sent to phasemeter by which the phase difference between two laser beams can be acquired.
To verify the periodic property of the nonlinear error caused by the stray light, the \ac{OPL} of the reference beam is increased linearly by using PZT-driven mirror.
The peak-to-peak value of the nonlinear error has been found to be 0.26 nm \cite{WOS:001254870400005}.

Regarding the laser interferometer for TianQin-2 mission, the prototype of the optical bench had been constructed and accomplished space-compatibility tests, like periodic vibrations, random vibrations and thermal cycle tests. The ground-based experiment shows that the noise level of the optical bench achieved 2 pm/Hz$^{1/2}$ @ 0.1 Hz and 6 pm/Hz$^{1/2}$ @ 10 mHz.

\subsubsection{Phasemeter and weak-light phase locking}\

The phasemeter for TianQin is being developed and has achieved a very high precision.
The current research has revealed that the measurement error of phasemeter is mainly caused by the phase sampling jitter that is closely related to the devices of \ac{ADC} and ultra-stable oscillator.
The former converts interference beat signal to digital signal, and the latter provides trigger signals to phasemeter and optical phase-locked loop.
The noises caused by the sampling jitter of \ac{ADC} and the noise of ultra-stable oscillator can be reduced by using the pilot-tone correction and the inter-satellite side-band modulation techniques, respectively. However, the low-frequency (0.1 mHz $\sim$ 0.1 Hz) noise of the phasemeter and weak-light phase locking control is still a challenging problem.

For low frequency noises of phasemeter, the major noise sources of phasemeter include thermal fluctuation of temperature-sensitive electrical device \cite{WOS:000358934400060}, inter-channel cross talk \cite{WOS:000358934400060,WOS:000758030200012}, and quantization error of numerical control oscillator \cite{Wand:2007bpa}.
We focus on discussion about the thermal fluctuation of temperature-sensitive electrical device in the following.

The phasemeter unit consists of a lot of analog and digital components and devices.
For analog components, like resistances and capacitances, their transfer functions can be obtained without much difficulty.
For more complicated devices, e.g. \ac{ADC} and multiplexers, it’s difficult to build up proper theoretical models for analyzing their thermal characteristics.
In this case, we apply temperature modulation to these complicated devices in order to calibrate the temperature coefficients \cite{WOS:001342123600001}.
Using such experiment, we can evaluate the thermal characteristics of temperature-sensitive components and devices used in the phasemeter and weak-light phase-locked loop.
Accordingly, we can determine the upper limits of temperature variations allowed for these temperature-sensitive components and devices.

\subsubsection{Telescope}\

{\it Coordinator: Lei Fan}

Telescope is another core payload of the laser interferometer system.
It functions as an afocal beam expander and operates simultaneously in transmit and receive modes.
Its primary purpose is to facilitate a precise length measurement between the optical benches (and ultimately the \acp{TM}) on widely separated \ac{S/C}.

The optical design for TianQin telescope adopts the off-axis quad-mirror optical structure. It takes a collimated beam from the optical bench, with a diameter of approximately 4 mm and transforms it into a collimated beam, with a diameter of about 300 mm, and a profile optimized to deliver power efficiently on-axis in the far field.
The latest design shows that the telescope has a magnification of 75, with M1 being parabolic, M2 hyperbolic, M3 a folding mirror, and M4 a freeform surface.
Within the $\pm$30 \textmu{}rad science field of view, wave front errors are better than RMS 0.0074 $\lambda$ ($\lambda$=1064nm).
Optical system design schemes where both M3 and M4 are freeform surfaces have also been attempted. Although higher specifications can be achieved in the design, the difficulty of manufacturing and alignment increases dramatically.
Therefore, they have not been chosen for prototype engineering.
The tolerance budget guarantees that there is over a 90\% probability that the telescope’s wave front error will be better than RMS 25 nm.
The spatial distances between M1 through M4 have been fully considered for compactness, structural support, alignment, and assembly requirements.
To fulfill the stability requirements, both the mirrors and supporting structures are made from low expansion glass, mainly due to its ultra-low coefficient of thermal expansion, ease of welding and bonding.
Under the premise of relaxing the wave front aberration, methods such as controlling the coupling aberration proportion and optimizing the beam waist radius have been proposed to suppress the non-geometric \ac{TTL} coupling noise \cite{Fan:24,Fan:2023cgl,Fan:2023xqi}.
In terms of coating, it has been found that metal coatings can effectively suppress the optical path noise caused by low-frequency temperature fluctuations \cite{Luo:2024aay}.

The optical path length stability of the telescope is a key technical indicator for the space-based \ac{GW} detection. In the observation band from 0.1 mHz to 1 Hz, the telescope must exhibit an optical path length stability of 0.4 pm/Hz$^{1/2}$.
We designed a detection scheme based on Pound-Drever-Hall technology and analyzed its noise requirement level \cite{Hai:2024cjl}.
Furthermore, an analysis indicates that when the detected telescope wave front aberration is better than 0.068 $\lambda$ ($\lambda$=1064 nm) with a probability of 98\%, the coupling efficiency of the off-axis resonant cavity can exceed 40\% \cite{Hai:24}.

\section{Summary}

The TianQin project aims to launch the space-based \ac{GW} detector, TianQin, around 2035. The project has been progressing smoothly following the ``0123" technology roadmap.
After the success of step ``0" and step ``1" missions, which have demonstrated the high orbit satellite laser ranging and the \ac{DFC} technology for TianQin, the step ``2" mission has been officially approved in 2021 and is expected to test the inter-satellite laser interferometry technology for TianQin using a pair of satellites in a couple of years.
Researches surrounding the final step, the construction and the launch of the \ac{GW} detection satellites, are also moving forward as expected.

\section*{Acknowledgments}

The work has been supported in part by the National Key Research and Development Program of China (Grant No. 2020YFC2200200, 2020YFC2200500, 2022YFC2204000, 2022YFC2204200, 2023YFC2206700),
the Guangdong Major Project of Basic and Applied Basic Research (Grant No. 2019B030302001),
the National Natural Science Foundation of China (Grants No. 12261131504, 11927812),
the Basic and Applied Basic Research Foundation of Guangdong Province (Grant Number 2024A1515010142),
the 111 Project (Grant No.B20062),
and the Fundamental Research Funds for the Central Universities, Sun Yat-sen University.
The work of VM and KP is supported by Russian Science Foundation grant No 23-42-00055.

\section*{References}
\bibliographystyle{unsrt}
\bibliography{ref_sec1,ref_sec2,ref_sec3,ref_sec4.1,ref_sec4.2}
\end{document}